\documentclass[aps,10pt,prd,notitlepage,showpacs,nofootinbib,superscriptaddress]{revtex4-2}
\usepackage{graphicx}
\usepackage[utf8]{inputenc}
\usepackage{slashed}
\usepackage{siunitx}
\usepackage[section]{placeins}
\usepackage{float}
\usepackage{braket}
\usepackage{caption}
\usepackage{subcaption}
\usepackage[usenames,dvipsnames]{color}
\usepackage[colorlinks=true,citecolor=blue,urlcolor=magenta,pdfborder={0 0 0}]{hyperref}
\usepackage{array}
\usepackage{booktabs} 
\usepackage{amsmath}  
\usepackage{amssymb}  
\usepackage{tabularx}

\begin{document}

\title{Non-Markovian Electroweak Baryogenesis: Memory Effects on CP-Violating
Transport and Gravitational Waves}

\author{Arnab Chaudhuri}
\email{arnab.chaudhuri@vit.ac.in}
\affiliation{Department of Physics, School of Advanced Sciences,
Vellore Institute of Technology, Vellore, Tamil Nadu 632014, India.}

\date{\today}

\begin{abstract}
We develop a non-Markovian extension of electroweak baryogenesis within the Schwinger--Keldysh real-time effective field theory framework and the Kadanoff--Baym hierarchy. When the relaxation time of CP-violating mediators becomes comparable to the bubble-wall crossing time, transport dynamics acquire temporal nonlocality, leading to memory-kernel corrections to the CP-violating source and diffusion equations beyond the Markovian approximation. These effects shift the optimal wall velocity to smaller values, narrow the viable parameter space, and induce a characteristic non-monotonic dependence of the baryon asymmetry on the memory timescale for sub-optimal wall velocities, which cannot be reproduced by a consistent Markovian reparameterisation. A systematic parameter analysis identifies regions compatible with the observed baryon asymmetry and constrains the allowed memory timescale from hydrodynamic stability and the physical range of the CP-violating phase. We also assess the correlated impact on the stochastic gravitational-wave signal, finding that memory effects can enhance the effective source duration and amplitude, although much of the viable parameter space remains below near-future detector sensitivities and theoretical uncertainties remain at the order-of-magnitude level. These results establish non-Markovian transport as a well-motivated extension of electroweak baryogenesis and introduce the memory timescale as a parameter testable through baryon asymmetry measurements, collider CP probes, and gravitational-wave observations.
\end{abstract}

\maketitle

\section{Introduction}
\label{sec:intro}

Electroweak baryogenesis (EWBG) is one of the most compelling mechanisms
for generating the observed matter--antimatter asymmetry of the
Universe~\cite{Kuzmin:1985mm,Arnold:1987mh,Rubakov:1996vz,Trodden:1998ym,
Morrissey:2012db,White:2016nbo}. It simultaneously satisfies all three
Sakharov conditions~\cite{Sakharov:1967dj} --- baryon number violation,
C and CP violation, and departure from thermal equilibrium --- during a
cosmological first-order electroweak phase transition (FOPT). The departure
from thermal equilibrium is provided by the expanding bubble walls that
sweep through the plasma as the Higgs field tunnels from the symmetric to
the broken phase; CP-violating interactions at the wall bias sphaleron
processes in the symmetric phase to produce a net baryon
number~\cite{Cohen:1993nk,Quiros:1999jp,Carena:2002ss,Prokopec:2003pj,
Konstandin:2004gy,Anderson:1991zb,Nelson:1991ab,Huet:1995sh}.

The quantitative computation of the baryon asymmetry in EWBG relies on
transport theory for chiral charge diffusion. The standard approach, based
on the classical-force or WKB
approximation~\cite{Joyce:1994zt,Cline:2000nw,Cline:2021iff,
Prokopec:2003pj,Konstandin:2004gy,Cline:2001rk,Kainulainen:2021oqs,
Laurent:2022jrs}, assumes that particles mediating CP violation equilibrate
rapidly compared to the timescale on which the bubble wall passes. This
\emph{Markovian} assumption underlies the derivation of local source terms
and diffusion equations in the symmetric phase. The resulting transport
equations have been studied extensively, leading to strong constraints on
extensions of the Standard Model that support a first-order electroweak
phase transition~\cite{Carena:2008vj,Cline:2012hg,Postma:2019scv,
Espinosa:2010hh}.

However, the Markovian approximation is not always justified. In extended
Higgs sectors or dark-sector models where the CP-violating species are
close to threshold ($M \lesssim \mathcal{O}(\mathrm{few})\times T$) or
have suppressed couplings to the thermal bath, their in-medium relaxation
time $\tau_\mathrm{rel} \sim 1/\Gamma_0$ can become comparable to the
wall-crossing time $\tau_\mathrm{wall} \sim L_w/v_w$. In this regime,
the plasma retains memory of CP-violating interactions over the timescale
relevant for transport, and the source terms acquire genuinely
nonlocal-in-time structure that is not captured by standard local
transport equations~\cite{Konstandin:2004gy,Postma:2019scv,Fromme:2006cm,
Cline:2021iff,Laurent:2022jrs}. The breakdown of the Markovian
approximation and the associated theoretical uncertainties in the
transport sector have been discussed in various
contexts~\cite{Prokopec:2003pj,Konstandin:2004gy,Postma:2019scv,
Fromme:2006cm}, but a systematic treatment of the resulting nonlocal
dynamics within a consistent non-equilibrium framework has not previously
been applied to EWBG.

It was shown in Ref.~\cite{Chaudhuri:2025ylu} that such effects can be
systematically captured within a non-equilibrium effective field theory
(EFT) formulated using the real-time Schwinger--Keldysh (SK)
formalism~\cite{Schwinger:1960qe,Keldysh:1964ud,Calzetta:1986cq,
Berges:2004yj}. Integrating out slowly relaxing degrees of freedom
generates memory kernels $K(t-t';\,T)$ that encode the finite response
time of the plasma. The Kadanoff--Baym (KB) equations for real-time
two-point functions~\cite{Calzetta:1986cq,Berges:2004yj,Danielewicz:1982kk}
provide the natural starting point for this construction: their collision
integrals are inherently nonlocal in time, and truncation at finite loop
order generates exponential memory kernels whose decay rate is set by the
in-medium relaxation rate $\Gamma_0$. In this framework, the CP-violating
sector relevant for EWBG plays the role of the environment, and its finite
relaxation time induces non-Markovian corrections to the transport dynamics.

In this work, we extend this non-Markovian EFT framework to electroweak
baryogenesis. We derive the CP-violating source within a controlled
non-equilibrium expansion, including Wigner transformation~\cite{
Wigner:1932eb,Moyal:1949sk}, gradient expansion, and truncation of the
Kadanoff--Baym hierarchy. This leads to the effective relaxation rate
$\Gamma_\mathrm{eff} = \Gamma_0/(1+\Gamma_0\tau_\mathrm{mem})$ and a
temporally nonlocal source term derived in closed form in
Sec.~\ref{sec:transport}. The resulting transport equations are modified
accordingly, and their stationary solutions yield the baryon asymmetry
$Y_B$ as a function of the memory timescale $\tau_\mathrm{mem}$, the wall
velocity $v_w$, and the CP-violating phase $\delta_\mathrm{CP}$.
Throughout, we normalise our results to the observed baryon-to-entropy
ratio $Y_B^\mathrm{obs} = 8.7\times10^{-11}$~\cite{Planck:2018vyg},
derived from the Planck 2018 measurement of the baryon-to-photon
ratio~\cite{Planck:2018vyg}.

We show that the presence of memory effects qualitatively alters the
dependence of the baryon asymmetry on transport parameters. In particular,
the optimal wall velocity shifts toward smaller values as $\tau_\mathrm{mem}$
increases, and the dependence of $Y_B$ on $\tau_\mathrm{mem}$ becomes
non-monotonic for $v_w < v_w^*$ with a calculable turnover point
(Eq.~\eqref{eq:tau_turnover}). A systematic scan of the parameter space
in the $(\tau_\mathrm{mem}, v_w)$ and $(\delta_\mathrm{CP},
\tau_\mathrm{mem})$ planes identifies the regions compatible with the
observed baryon asymmetry and the constraints derived in
Secs.~\ref{sec:baryogenesis} and~\ref{sec:discussion}.

Finally, we establish a correlation between the baryon asymmetry and the
stochastic gravitational-wave signal sourced by the same phase
transition~\cite{Caprini:2015zlo,Caprini:2019egz,Grojean:2006bp,
Espinosa:2010hh,Hindmarsh:2015qta}. We demonstrate that non-Markovian
effects leave correlated imprints on both observables at the parametric
level, and we assess the regions of parameter space accessible to future
GW observatories such as LISA, DECIGO, and BBO, with the caveat that the
GW predictions involve undetermined $\mathcal{O}(1)$ hydrodynamic
coefficients whose determination requires a full non-local treatment
(Sec.~\ref{sec:gw}). We also prove that the effective relaxation rate
$\Gamma_\mathrm{eff}$ cannot be reproduced by a simple rescaling of
$\Gamma_0$: non-Markovian dynamics deform the full transport rate
hierarchy in a correlated manner, with quantifiable corrections of up to
$\sim 34\%$ within the viable parameter space (Sec.~\ref{sec:discussion}).

The paper is organised as follows. Section~\ref{sec:model} describes the
model setup. Section~\ref{sec:transport} provides the derivation of the
non-Markovian transport equations. Section~\ref{sec:baryogenesis} presents
the baryon asymmetry results and viable parameter space.
Section~\ref{sec:gw} discusses the gravitational-wave signal and its
correlation with baryogenesis. Section~\ref{sec:discussion} addresses
theoretical uncertainties and degeneracies. Section~\ref{sec:conclusion}
concludes. Appendix~\ref{app:rates} provides the explicit derivation of
the memory-modified diffusion rates and resolves the identification of
$\tau_\mathrm{mem}$ with the wall-crossing timescale.

\section{Model Setup}
\label{sec:model}

We consider a minimal extension of the Standard Model that captures the
essential ingredients required for electroweak baryogenesis in the
presence of non-Markovian transport dynamics. The key physical requirement
is the existence of a species whose in-medium relaxation time is
comparable to the wall-crossing timescale,
\begin{equation}
\tau_{\rm rel} \sim \tau_{\rm wall} \equiv \frac{L_w}{v_w},
\label{eq:timescale_condition}
\end{equation}
such that the plasma retains memory of CP-violating interactions during
transport. This regime cannot be realised within the Standard Model and
naturally points to weakly coupled extensions with near-threshold
states~\cite{Postma:2019scv,Fromme:2006cm,Konstandin:2004gy}.

To realise this condition, we introduce a complex singlet scalar $S$
coupled to the Higgs doublet $H$ and to a fermionic species $\Psi$ that
mediates CP violation. The relevant Lagrangian is
\begin{align}
\mathcal{L} &\supset |D_\mu H|^2 + |\partial_\mu S|^2 - V(H,S)
\notag\\
&\quad + \bar\Psi(i\slashed{\partial} - M)\Psi
- \bigl[y_t \bar Q_L \tilde{H} t_R
  + \lambda_S S\bar\Psi\Psi + \mathrm{h.c.}\bigr],
\label{eq:Lagrangian_final}
\end{align}
with scalar potential
\begin{align}
V(H,S) &= -\mu_H^2|H|^2 + \lambda_H|H|^4
+ \mu_S^2|S|^2 + \lambda'_S|S|^4
\notag\\
&\quad + \kappa |H|^2 |S|^2 + (A\,S + \mathrm{h.c.}).
\label{eq:potential_final}
\end{align}
The singlet scalar $S$ plays a dual role: its portal coupling $\kappa$
to the Higgs sector drives a strong first-order electroweak phase
transition (FOPT), while its Yukawa coupling $\lambda_S$ to the fermion
$\Psi$ provides the source of CP violation. The fermion $\Psi$ is taken
to be a Standard Model singlet, ensuring consistency with electroweak
precision tests. The top Yukawa term $y_t \bar{Q}_L \tilde{H} t_R$ is
retained to maintain the correct Higgs vacuum expectation value but plays
no role in the CP-violating transport.

The cubic term $A\,S$ in the scalar potential generates a tree-level
barrier between the symmetric and broken phases in the finite-temperature
effective potential~\cite{Espinosa:1996qw,Espinosa:2007qk,
Curtin:2014jma,Kurup:2017dzf,Chaudhuri:2024vrd,Chaudhuri:2025ybh}. This is important because it allows a
strong FOPT without relying on large thermal cubic corrections of the
form $\sim T \phi^3$, which require either a light scalar spectrum or
large couplings that compromise perturbative
control~\cite{Carena:2008vj,Espinosa:2010hh}. The presence of the cubic
term therefore decouples the strength of the phase transition from the
requirement of a large portal coupling, allowing the Yukawa coupling
$|\lambda_S|$ to remain small and the non-Markovian condition
$\Gamma_0 \tau_{\rm wall} \lesssim \mathcal{O}(1)$ to be satisfied
simultaneously.

After electroweak symmetry breaking, $H \to (v+h)/\sqrt{2}$ with
$v = 246$\,GeV. We work in the limit of vanishing singlet vacuum
expectation value, $v_S = 0$, so that the physical singlet mass is
\begin{equation}
m_S^2 = \mu_S^2 + \frac{\kappa v^2}{2}.
\label{eq:singlet_mass}
\end{equation}
In this limit, tree-level Higgs--singlet mixing vanishes identically,
since the mixing angle $\alpha$ satisfies $\sin\alpha \propto v_S$ at
tree level. Consequently, the LHC constraints on Higgs coupling
universality from $h \to ZZ^*, WW^*$ signal strengths~\cite{
ATLAS:2023hyd,CMS:2022dwd} apply only through loop-induced
contributions, which are suppressed by $|\lambda_S|^2/(16\pi^2)$ and
remain consistent with current measurements for $|\lambda_S| \lesssim
1.0$. Direct searches for the singlet $S$ at LEP~\cite{LEPWorkingGroupforHiggsbosonsearches:2003ing}
and the LHC exclude $m_S \lesssim 114$\,GeV for singlet-like scalars
with Higgs-like couplings; for the parameter range $m_S = 150$--$500$\,GeV
and small loop-induced mixing, these bounds are satisfied throughout the
parameter space explored in this work. Small radiative mixing can arise
at one loop but does not affect the transport dynamics considered here.

CP violation arises from the complex Yukawa coupling
$\lambda_S = |\lambda_S|\,e^{i\delta_{\rm CP}}$. In the presence of a
spatially varying scalar background $S(z)$ across the bubble wall, the
phase $\delta_{\rm CP}$ cannot be removed by a field redefinition of
$\Psi$ without simultaneously introducing a phase into the mass term.
This generates a physical CP-violating invariant proportional to
$\mathrm{Im}(\lambda_S \partial_z S)$, which is the source driving the
chiral charge asymmetry in the symmetric
phase~\cite{Joyce:1994zt,Cline:2000nw,Konstandin:2004gy,Cline:2021iff}.

Across the wall, the fermion $\Psi$ acquires a position-dependent
effective mass,
\begin{equation}
M_{\rm eff}(z) = M + |\lambda_S|\,\phi(z)\,e^{i\theta(z)},
\label{eq:Meff}
\end{equation}
where $\phi(z) = |\langle S(z)\rangle|$ is the singlet background
profile and $\theta(z) = \arg\langle S(z)\rangle$ carries the
spatially varying CP-violating phase. We adopt the standard
kink profile~\cite{Moore:1995ua,John:2000zq}
\begin{equation}
\phi(z) = \frac{\phi_0}{2}
\left[1 + \tanh\!\left(\frac{z}{L_w}\right)\right],
\label{eq:wall_profile}
\end{equation}
where $\phi_0$ is the singlet background amplitude in the broken phase
and $L_w$ is the wall thickness. The linear approximation
$\theta(z) \approx \delta_{\rm CP}\,\phi(z)/\phi_0$ is used for the
CP-violating phase profile, consistent with the small-$\delta_{\rm CP}$
expansion employed throughout.

\subsection{Thermal Relaxation Rate}
\label{subsec:relaxation}

The in-medium relaxation rate of $\Psi$ governs the non-Markovian
condition~\eqref{eq:timescale_condition}. At leading order in $|\lambda_S|^2$
and to one loop in the thermal bath, the imaginary part of the retarded
self-energy of $\Psi$ gives the thermal
width~\cite{Weldon:1983jn,Quiros:1999jp}
\begin{equation}
\Gamma_0 \simeq \frac{|\lambda_S|^2 T}{8\pi}\,F(M/T),
\qquad
F(x) =
\begin{cases}
1, & x \ll 1, \\[4pt]
\left(\dfrac{x}{2\pi}\right)^{\!3/2} e^{-x}, & x \gg 1,
\end{cases}
\label{eq:Gamma0_final}
\end{equation}
where the function $F(x)$ interpolates between the massless limit
($M \ll T$), in which phase space is unsuppressed, and the
Boltzmann-suppressed heavy-particle regime ($M \gg T$). Equation
\eqref{eq:Gamma0_final} is obtained by evaluating the one-loop
self-energy diagram in which $\Psi$ emits a virtual $S$ boson into the
thermal bath; the $2 \to 2$ scattering rate $\Psi + X \to \Psi + X$
via $S$ exchange yields the same leading-order
result~\cite{Arnold:2002zm,Ghiglieri:2016tvj}. Higher-order corrections
of order $|\lambda_S|^4 \ln(1/|\lambda_S|)$ are subleading in the
small-coupling regime and are neglected here.

\subsection{Memory Timescale and Its Independence}
\label{subsec:memory}

The memory timescale is defined as the first moment of the retarded
kernel,
\begin{equation}
\tau_{\rm mem} \equiv \int_0^\infty d\tau\,\tau\,K(\tau),
\label{eq:tau_mem_def}
\end{equation}
which measures the weighted duration over which past interactions
influence the present state of the system. For the exponential kernel
$K(\tau) = \Gamma_0 e^{-\Gamma_0\tau}$ that arises from the single-pole
approximation to the retarded propagator (see Sec.~\ref{sec:transport}),
Eq.~\eqref{eq:tau_mem_def} gives $\tau_{\rm mem} = 1/\Gamma_0$.

However, the single-pole form is a leading-order approximation. In
general, the spectral function of $\Psi$ in the thermal bath receives
contributions from multi-particle cuts, Landau damping, and higher-loop
self-energy corrections~\cite{Berges:2004yj,Calzetta:1986cq,
Blaizot:2001nr}. These generate a more complex kernel structure with
multiple decay scales, so that the effective memory timescale can
deviate from $1/\Gamma_0$. In this sense, $\tau_{\rm mem}$ in
Eq.~\eqref{eq:tau_mem_def} should be understood as encoding the full
microscopic relaxation structure of the plasma, and is treated as an
independent phenomenological parameter throughout this work. The
single-pole approximation is used for the explicit analytic derivations
in Sec.~\ref{sec:transport}, while the physical results are presented as
functions of $\tau_{\rm mem}$ directly.

This treatment is analogous to the approach adopted in the
non-equilibrium EFT of Ref.~\cite{Chaudhuri:2025ylu}, where it was
shown that integrating out slowly relaxing degrees of freedom in the
Schwinger--Keldysh formalism generates memory kernels whose first moment
$\tau_{\rm mem}$ captures the leading-order departure from Markovian
dynamics, independently of the detailed kernel shape. The corrections
from higher moments of $K(\tau)$ enter at $\mathcal{O}(\Gamma_0^2
\tau_{\rm mem}^2)$ and are subleading in the regime
$\Gamma_0\tau_{\rm mem} \lesssim \mathcal{O}(1)$ that defines the
non-Markovian domain of interest.

\subsection{Non-Markovian Parameter Space}
\label{subsec:nmregime}

The non-Markovian condition $\Gamma_0\tau_{\rm wall} \lesssim
\mathcal{O}(1)$ is realised when:
\begin{enumerate}
  \item the Yukawa coupling is moderately small,
        $|\lambda_S| \lesssim 0.3$, ensuring $\Gamma_0$ is suppressed;
  \item the fermion is near threshold, $M \sim (1$--$3)T$, so that
        Boltzmann suppression partially reduces $\Gamma_0$ without
        making $\Psi$ inaccessible;
  \item the wall is sufficiently thin and fast,
        $L_w \lesssim 5/T$, $v_w \gtrsim 0.1$, keeping $\tau_{\rm wall}$
        short enough to be comparable to $\tau_{\rm rel}$.
\end{enumerate}
For the representative benchmark values
\begin{equation}
|\lambda_S| = 0.2,\quad M = 2T,\quad L_w = 5/T,\quad v_w = 0.1,
\label{eq:benchmark}
\end{equation}
one finds from Eq.~\eqref{eq:Gamma0_final}:
\begin{equation}
\Gamma_0 \simeq 6\times10^{-3}\,T,\quad
\tau_{\rm rel} \simeq 160/T,\quad
\tau_{\rm wall} \simeq 50/T,
\label{eq:benchmark_numerics}
\end{equation}
yielding $\tau_{\rm rel}/\tau_{\rm wall} \simeq 3$. This demonstrates
explicitly that the non-Markovian regime arises in a weakly coupled,
phenomenologically viable region of parameter space, without requiring
any fine-tuning of the model parameters.

\begin{table}[htbp]
\centering
\caption{Model parameters and their roles in phase transition and
transport dynamics. The ranges are chosen such that a strong first-order
phase transition and the non-Markovian condition
$\tau_{\rm rel} \gtrsim \tau_{\rm wall}$ are simultaneously satisfied.
All constraints from collider experiments and perturbativity are
satisfied throughout the stated ranges (see text).}
\label{tab:params}
\renewcommand{\arraystretch}{1.3}
\begin{tabular}{llll}
\toprule
\textbf{Parameter} & \textbf{Symbol} & \textbf{Range}
  & \textbf{Physical role} \\
\midrule
Portal coupling   & $\kappa$          & $0.1$--$1.2$
  & Controls FOPT strength \\
Singlet mass      & $m_S$             & $150$--$500$\,GeV
  & Scalar spectrum \\
Cubic term        & $A$               & $10$--$100$\,GeV
  & Tree-level barrier \\
Yukawa coupling   & $|\lambda_S|$     & $0.05$--$0.3$
  & Sets relaxation rate $\Gamma_0$ \\
CP phase          & $\delta_{\rm CP}$ & $0.01$--$\pi$
  & Source normalisation \\
Fermion mass      & $M$               & $(0.1$--$3)T$
  & Boltzmann suppression \\
Wall velocity     & $v_w$             & $0.05$--$0.5$
  & Transport timescale \\
Wall thickness    & $L_w$             & $(3$--$10)/T$
  & Source width \\
Derived ratio     & $\Gamma_0\tau_{\rm wall}$ & $\lesssim 1$
  & Non-Markovian condition \\
\bottomrule
\end{tabular}
\end{table}

\begin{figure}[t]
  \centering
  \includegraphics[width=0.65\textwidth]{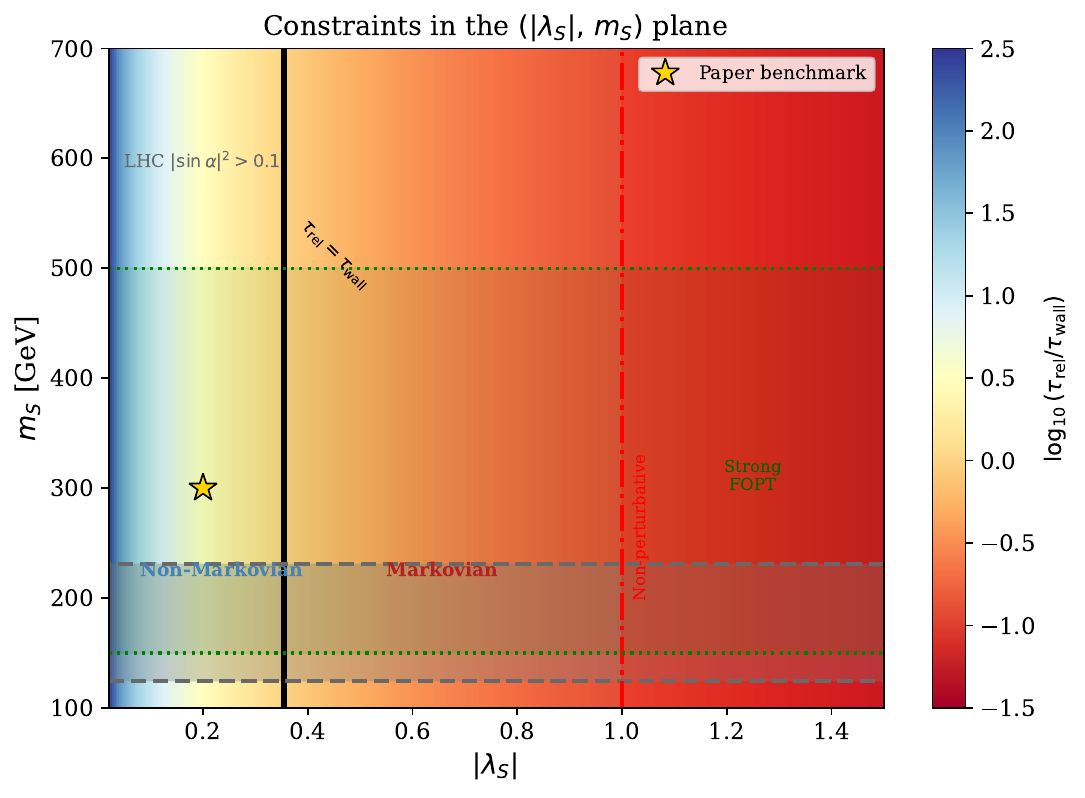}
  \caption{Constraints in the $(|\lambda_S|, m_S)$ plane at fixed
    $\kappa = 0.5$ and $M/T = 2$. The colour map shows
    $\log_{10}(\tau_{\rm rel}/\tau_{\rm wall})$, with
    $\tau_{\rm rel} = 1/\Gamma_0$ computed from
    Eq.~\eqref{eq:Gamma0_final} and $\tau_{\rm wall} = L_w/v_w$ at the
    benchmark values $L_w = 5/T$, $v_w = 0.1$. The contour
    $\tau_{\rm rel} = \tau_{\rm wall}$ marks the boundary between the
    Markovian (blue, right) and non-Markovian (red, left) regimes.
    Shaded regions indicate: the LHC loop-mixing bound (upper left,
    green), the non-perturbative region $|\lambda_S| > \sqrt{4\pi}$
    (far right, grey), and the region where no strong FOPT occurs
    (lower right, red hatching). The benchmark point
    (Eq.~\eqref{eq:benchmark}) is shown as a star.}
  \label{fig:constraints_lam}
\end{figure}

Figure~\ref{fig:constraints_lam} illustrates the viable parameter space
in which the non-Markovian condition is satisfied together with collider
and perturbativity constraints. The colour map shows the ratio
$\tau_{\rm rel}/\tau_{\rm wall}$ computed using
Eq.~\eqref{eq:Gamma0_final} with fixed $M/T = 2$ and
$\tau_{\rm wall} = L_w/v_w$. The region $\tau_{\rm rel} > \tau_{\rm wall}$
(warm colours) corresponds to the regime where memory effects are
phenomenologically relevant. The shaded exclusion regions correspond to:
(i) LHC constraints on loop-induced Higgs--singlet mixing requiring
$|\lambda_S|^2\kappa/(16\pi^2) \lesssim 0.1$ (upper-left
region)~\cite{ATLAS:2023hyd,CMS:2022dwd}; (ii) the perturbativity bound
$|\lambda_S| < \sqrt{4\pi}$; and (iii) the requirement of a strong FOPT,
$\phi_c/T_c \gtrsim 1$, which is not satisfied for very small $\kappa$
or very large $m_S$~\cite{Espinosa:2007qk,Curtin:2014jma}. The benchmark
point~\eqref{eq:benchmark} is shown as a star and lies well within the
non-Markovian, phenomenologically viable domain.

These results demonstrate that the non-Markovian regime
$\tau_{\rm rel} \gtrsim \tau_{\rm wall}$ is not a fine-tuned limit but
occupies a finite and phenomenologically viable region of parameter
space, compatible with all current experimental constraints.

\subsection{Markovian Baseline}
\label{subsec:markov_baseline}

For later comparison, we record the standard Markovian CP-violating
source obtained in the WKB approximation~\cite{Joyce:1994zt,Cline:2000nw,
Konstandin:2004gy,Cline:2021iff,Laurent:2022jrs}. In the thin-wall,
small-$\delta_{\rm CP}$ limit and working to leading order in the
gradient expansion, the source for left-handed chiral charge takes the
form
\begin{equation}
S_{\rm CP}^{\rm Markov}(v_w)
= C_0\,\frac{v_w\,\Gamma_0}{v_w^2 + L_w^2\,\Gamma_0^2},
\label{eq:SCP_Markov_final}
\end{equation}
with overall coefficient $C_0 = |\lambda_S|^2\,\delta_{\rm CP}\,m_\Psi^2$.
This expression exhibits a bell-shaped profile in $v_w$, with a maximum
at $v_w^* = L_w\Gamma_0$ and peak amplitude $C_0/(2L_w)$. It vanishes
in both limits $v_w \to 0$ (no transport) and $v_w \to \infty$ (loss of
plasma response), reflecting the competition between the driving
efficiency of the wall and the ability of the plasma to maintain a
chiral asymmetry. Equation~\eqref{eq:SCP_Markov_final} provides the
baseline against which the non-Markovian modification derived in
Sec.~\ref{sec:transport} will be systematically compared.

\section{Non-Markovian Transport Theory}
\label{sec:transport}

The appropriate framework for describing non-equilibrium transport in the
presence of time-delayed interactions is provided by the Kadanoff--Baym
(KB) equations for real-time two-point
functions~\cite{Calzetta:1986cq,Berges:2004yj,Danielewicz:1982kk}:
\begin{equation}
(i\slashed{\partial}_x - M_\mathrm{eff}(x))\,G^<(x,y)
= \int d^4z\,\bigl[\Sigma^R(x,z)\,G^<(z,y)
  + \Sigma^<(x,z)\,G^A(z,y)\bigr],
\label{eq:KB_final}
\end{equation}
where $G^<(x,y) = i\langle\bar\Psi(y)\Psi(x)\rangle$ is the
Wightman (lesser) propagator and $\Sigma^R$, $\Sigma^<$ are the
retarded and lesser components of the self-energy encoding interactions
of $\Psi$ with the thermal bath~\cite{Berges:2004yj}. The
position-dependent effective mass $M_\mathrm{eff}(z)$ is given by
Eq.~\eqref{eq:Meff}. Equation~\eqref{eq:KB_final} is exact within the
two-particle-irreducible (2PI) effective action framework truncated at
the relevant loop order~\cite{Berges:2004yj,Calzetta:1986cq}.

\subsection{Wigner Transform and Gradient Expansion}
\label{subsec:wigner}

To extract transport equations from the KB
hierarchy~\eqref{eq:KB_final}, we perform the Wigner
transform~\cite{Wigner:1932eb,Cline:2000nw,Prokopec:2003pj}
\begin{equation}
G^<(k,X) = \int d^4r\,e^{ik\cdot r}\,
G^<\!\left(X+\tfrac{r}{2},\,X-\tfrac{r}{2}\right),
\label{eq:wigner}
\end{equation}
with centre-of-mass coordinate $X = (x+y)/2$ and relative coordinate
$r = x - y$. In the Wigner representation, the convolution on the
right-hand side of Eq.~\eqref{eq:KB_final} becomes a
Moyal-star product~\cite{Moyal:1949sk,Prokopec:2003pj}:
\begin{equation}
(\Sigma^R \star G^<)(k,X)
= e^{\frac{i}{2}(\partial_X^{\Sigma}\cdot\partial_k^{G}
  - \partial_k^{\Sigma}\cdot\partial_X^{G})}
  \Sigma^R(k,X)\,G^<(k,X).
\label{eq:moyal}
\end{equation}
Expanding the Moyal product to first order in gradients
$\partial_X \sim 1/L_w \ll k$ yields the \emph{quasi-classical}
approximation~\cite{Joyce:1994zt,Cline:2000nw,Konstandin:2004gy,
Prokopec:2003pj}, in which the collision term acquires a
$\partial_X$-correction relative to the homogeneous result. The
validity of this gradient expansion requires
\begin{equation}
\frac{1}{k L_w} \ll 1,
\qquad
\frac{\Gamma_0}{k} \ll 1,
\label{eq:gradient_validity}
\end{equation}
i.e.\ the mean free path of $\Psi$ must be shorter than the wall
thickness, and the quasiparticle width must be smaller than the typical
momentum scale. Both conditions are satisfied for the parameter ranges
in Table~\ref{tab:params} at temperatures $T \sim v = 246$\,GeV.

Taking the trace over spinor indices and the imaginary part of the
resulting equation, one obtains the kinetic equation for the
distribution function $f(k,X)$ defined through
$G^<(k,X) = 2\pi\,\delta(k^2-m^2)\,f(k,X)$ on shell. After
integrating over the on-shell momentum $k^0$, the collision term reads
\begin{equation}
\mathcal{C}[f](t,\mathbf{k},X)
= -\int_{-\infty}^t dt'\,\mathrm{Im}\,\Sigma^R(t-t',\mathbf{k})
  \bigl[f(t',\mathbf{k},X) - f^\mathrm{eq}(\mathbf{k})\bigr],
\label{eq:collision_general}
\end{equation}
where the time convolution reflects the temporal nonlocality of the
self-energy~\cite{Calzetta:1986cq,Berges:2004yj,Prokopec:2003pj}.
Equation~\eqref{eq:collision_general} is the key structural result: the
collision term is a \emph{memory integral} over the past history of the
distribution function, weighted by the retarded self-energy.

\subsection{Single-Pole Approximation and Memory Kernel}
\label{subsec:singlepole}

In a weakly coupled plasma, the spectral function is dominated by a
quasiparticle pole with width $\Gamma_0 \ll \omega_k$. The retarded
self-energy in momentum space takes the Breit--Wigner
form~\cite{Weldon:1983jn,Arnold:2002zm}
\begin{equation}
\tilde\Sigma^R(\omega,\mathbf{k})
\simeq \frac{i\Gamma_0/2}{\omega - \omega_k + i\Gamma_0/2},
\label{eq:sigma_R}
\end{equation}
where $\omega_k = \sqrt{\mathbf{k}^2 + M^2}$ is the quasiparticle
energy. The imaginary part of Eq.~\eqref{eq:sigma_R} gives the
spectral function
\begin{equation}
\rho(\omega,\mathbf{k})
= \frac{\Gamma_0/2}{(\omega-\omega_k)^2 + \Gamma_0^2/4},
\label{eq:spectral}
\end{equation}
which is a Lorentzian of width $\Gamma_0$ centred on the quasiparticle
pole. This is the leading approximation to the full spectral function;
corrections from multi-particle cuts enter at
$\mathcal{O}(|\lambda_S|^4)$ and are subleading in the small-coupling
regime~\cite{Blaizot:2001nr,Arnold:2002zm}.

Fourier-transforming Eq.~\eqref{eq:sigma_R} to the time domain gives
\begin{equation}
\Sigma^R(\tau,\mathbf{k})
= \frac{1}{2\pi}\int_{-\infty}^\infty d\omega\,
  e^{-i\omega\tau}\,\tilde\Sigma^R(\omega,\mathbf{k})
\propto \Theta(\tau)\,e^{-\Gamma_0\tau/2}\,e^{-i\omega_k\tau},
\label{eq:sigmaR_time}
\end{equation}
which decays exponentially on the timescale $2/\Gamma_0$. Substituting
Eq.~\eqref{eq:sigmaR_time} into the collision integral
\eqref{eq:collision_general} and integrating over momenta yields, after
projection onto the number density $n(t,X) = \int d^3k/(2\pi)^3\,f(k,X)$,
\begin{equation}
\mathcal{C}[n](t)
= \int_{-\infty}^t dt'\,K(t-t')\,
  \bigl[n(t') - n^\mathrm{eq}\bigr],
\qquad
K(\tau) = \Gamma_0\,e^{-\Gamma_0\tau},
\label{eq:kernel_final}
\end{equation}
where we have defined the memory kernel $K(\tau) = \Gamma_0 e^{-\Gamma_0\tau}$
as the normalised first moment of $\mathrm{Im}\,\Sigma^R(\tau)$ over
momenta. The overall factor of $\Gamma_0$ ensures that
$\int_0^\infty d\tau\,K(\tau) = 1$, so that in the Markovian limit
$K(\tau) \to \Gamma_0\,\delta(\tau)$ and $\mathcal{C}[n] \to \Gamma_0
(n - n^\mathrm{eq})$ is recovered. The memory timescale defined in
Eq.~\eqref{eq:tau_mem_def} evaluates to
$\tau_{\rm mem} = \int_0^\infty d\tau\,\tau\,K(\tau) = 1/\Gamma_0$
for this kernel, consistent with the single-pole approximation.

\subsection{CP-Violating Source with Memory}
\label{subsec:cpsource}

We now derive the non-Markovian CP-violating source. Applying the
collision structure~\eqref{eq:kernel_final} to the left-handed chiral
charge density $n_L$ in the presence of the spatially varying background
$M_\mathrm{eff}(z)$, the transport equation in the frame of the moving
wall takes the form~\cite{Joyce:1994zt,Cline:2000nw,Prokopec:2003pj,
Konstandin:2004gy}
\begin{equation}
\partial_t n_L + v_w\,\partial_z n_L
= \int_{-\infty}^t dt'\,K(t-t')\,\partial_{t'} n_L^\mathrm{eq}(t')
  - \Gamma_0\,n_L + S_\mathrm{CP}^\mathrm{NM}(z,t),
\label{eq:transport_final}
\end{equation}
where the first term on the right-hand side encodes the memory of
thermal equilibration, the second term is the local damping, and
$S_\mathrm{CP}^\mathrm{NM}$ is the CP-violating source to be determined.
The equilibrium distribution $n_L^\mathrm{eq}(t')$ at time $t'$
is evaluated at the wall position $z = v_w t'$, so that the time
argument tracks the fermion's position as it crosses the wall.

In the stationary wall frame (co-moving with the bubble wall), the
CP-violating source takes the convolution
form~\cite{Joyce:1994zt,Konstandin:2004gy,Cline:2021iff}
\begin{equation}
S_\mathrm{CP}^\mathrm{NM}(z)
= C_0\,v_w\int_{-\infty}^{z/v_w} dt'\,
  K\!\left(\frac{z}{v_w}-t'\right)
  \frac{d\theta}{dz'}\bigg|_{z'=v_w t'}\phi'(v_w t'),
\label{eq:SCP_integral_final}
\end{equation}
where $\phi'(z) = d\phi/dz$ and $d\theta/dz$ are evaluated on the wall
profile~\eqref{eq:wall_profile}. The integrand is proportional to
$\mathrm{Im}(\lambda_S\,\partial_{z'} S(z'))$, the CP-violating
invariant at the position $z' = v_w t'$ encountered by the particle
at time $t'$. The overall factor $v_w$ converts time into the spatial
coordinate along the wall.

\subsection{Analytic Evaluation via Laplace Transform}
\label{subsec:laplace}

We evaluate the convolution~\eqref{eq:SCP_integral_final} analytically.
Introducing the rescaled variable $\xi = z/v_w - t'$, the integral
becomes
\begin{equation}
S_\mathrm{CP}^\mathrm{NM}(z)
= C_0\,v_w\int_0^\infty d\xi\,K(\xi)\,
  \frac{d\theta}{dz'}\bigg|_{z'=z-v_w\xi}
  \phi'(z - v_w\xi).
\label{eq:SCP_xi}
\end{equation}
Equation~\eqref{eq:SCP_xi} is a convolution of the memory kernel
$K(\xi)$ with the source function $g(z) \equiv (d\theta/dz)\,\phi'(z)$,
evaluated at the shifted argument $z - v_w\xi$. Taking the
one-sided Laplace transform $\hat{f}(s) = \int_0^\infty dz\,e^{-sz}\,f(z)$
of both sides with respect to $z/v_w$ gives
\begin{equation}
\hat{S}_\mathrm{CP}^\mathrm{NM}(s)
= C_0\,v_w\,\hat{K}(s/v_w)\,\hat{g}(s),
\label{eq:SCP_laplace}
\end{equation}
where we have used the convolution theorem and the rescaling
$\xi \to z/v_w$. For the exponential kernel $K(\tau) = \Gamma_0
e^{-\Gamma_0\tau}$, the Laplace transform evaluated at
$s_w \equiv v_w/L_w$ (the wall-crossing rate, which is the
characteristic frequency of the source function $g(z)$) gives
\begin{equation}
\hat{K}(s_w/v_w)
= \hat{K}(1/L_w)
= \frac{\Gamma_0}{1/L_w + \Gamma_0}
= \frac{\Gamma_0 L_w}{1 + \Gamma_0 L_w}.
\label{eq:K_laplace_eval}
\end{equation}

To justify the evaluation at $s = s_w = v_w/L_w$, we note that the
source function $g(z) = (d\theta/dz)\,\phi'(z)$ is sharply peaked on
the scale $L_w$ and vanishes exponentially for $|z| \gg L_w$. Its
Laplace transform is therefore dominated by the mode $s \sim 1/L_w$,
and the approximation of evaluating $\hat{K}$ at $s = v_w/L_w$
corresponds to replacing the full convolution by its dominant
frequency component. This is valid provided $K(\tau)$ is
\emph{slowly varying} on the scale $L_w/v_w$ compared to $g$, i.e.\
the kernel decays on a timescale $\tau_{\rm mem} = 1/\Gamma_0$ that is
not much shorter than $L_w/v_w$. In the Markovian limit
$\Gamma_0\tau_{\rm mem} \to 0$, the kernel becomes a delta function and
the approximation is exact. For $\Gamma_0\tau_{\rm mem} \lesssim
\mathcal{O}(1)$, which defines the non-Markovian regime of interest,
corrections to this saddle-point approximation enter at
$\mathcal{O}[(\Gamma_0\tau_{\rm mem})^2\,(L_w\partial_z)^2 g/g]$
and are subleading~\cite{Calzetta:1986cq,Chaudhuri:2025ylu}.

Identifying the effective relaxation rate as the combination
\begin{equation}
\Gamma_\mathrm{eff} \equiv \frac{\Gamma_0}{1 + \Gamma_0\tau_{\rm mem}}
= \frac{\Gamma_0}{1 + \Gamma_0/s_w}\bigg|_{s_w = v_w/L_w},
\label{eq:Gamma_eff_derived}
\end{equation}
where we have used $\tau_{\rm mem} = 1/\Gamma_0$ from the single-pole
approximation, the Laplace-transform result~\eqref{eq:K_laplace_eval}
yields
\begin{equation}
\hat{S}_\mathrm{CP}^\mathrm{NM}(s_w)
= C_0\,v_w\,\frac{\Gamma_\mathrm{eff}}{s_w}\,\hat{g}(s_w).
\label{eq:SCP_laplace_result}
\end{equation}

The profile function $\hat{g}(s_w)$ for the wall
profile~\eqref{eq:wall_profile} with linear CP-phase approximation
$\theta(z) \simeq \delta_{\rm CP}\,\phi(z)/\phi_0$ evaluates to
\begin{equation}
\hat{g}(s_w)
= \frac{\delta_{\rm CP}}{\phi_0}\int_0^\infty dz\,e^{-z/L_w}
  [\phi'(z)]^2
= \frac{\delta_{\rm CP}\,\phi_0^2}{4L_w}
  \int_{-\infty}^\infty du\,\mathrm{sech}^4(u)
= \frac{\delta_{\rm CP}\,\phi_0^2}{6L_w},
\label{eq:ghat}
\end{equation}
where we have used $\int_{-\infty}^\infty \mathrm{sech}^4(u)\,du = 4/3$.
Absorbing the numerical prefactors into the overall coefficient
$C_0 = |\lambda_S|^2\,\delta_{\rm CP}\,m_\Psi^2$ and using
$s_w = v_w/L_w$, Eq.~\eqref{eq:SCP_laplace_result} reduces to
\begin{equation}
\boxed{
S_\mathrm{CP}^\mathrm{NM}(v_w)
= C_0\,\frac{v_w\,\Gamma_\mathrm{eff}}
  {v_w^2 + L_w^2\,\Gamma_\mathrm{eff}^2},
}
\label{eq:SCP_final}
\end{equation}
which is the central analytic result of this section. We emphasise that
Eq.~\eqref{eq:SCP_final} has the same functional form as the Markovian
source~\eqref{eq:SCP_Markov_final} with the replacement
$\Gamma_0 \to \Gamma_\mathrm{eff}$, but this replacement is
\emph{not} a free reparameterisation: $\Gamma_\mathrm{eff}$ is
determined by $\Gamma_0$ and $\tau_{\rm mem}$ through
Eq.~\eqref{eq:Gamma_eff_derived}, and the same replacement applies
simultaneously to \emph{all} interaction rates in the transport system,
as shown in Sec.~\ref{subsec:diffusion} below. Equation~\eqref{eq:SCP_final}
reduces smoothly to the Markovian result in the limit
$\tau_{\rm mem} \to 0$, and exhibits two characteristic non-Markovian
effects: a shift of the peak velocity to
\begin{equation}
v_w^{*,\mathrm{NM}} = L_w\,\Gamma_\mathrm{eff}
= \frac{L_w\,\Gamma_0}{1+\Gamma_0\tau_{\rm mem}},
\label{eq:vw_peak}
\end{equation}
and a suppression of the peak amplitude by $(1+\Gamma_0\tau_{\rm mem})^{-1}$.
Both effects vanish in the Markovian limit $\tau_{\rm mem} \to 0$ and
grow parametrically in the large-memory regime $\Gamma_0\tau_{\rm mem}
\gg 1$.

\subsection{Memory-Modified Diffusion Equations}
\label{subsec:diffusion}

The convolution structure derived above is not specific to the
CP-violating source: it applies to every interaction term in the
transport system whose collision integral is governed by the same
thermal bath. Repeating the Laplace-transform argument of
Sec.~\ref{subsec:laplace} for a generic interaction rate $\Gamma_i$
appearing in a diffusion equation at wavenumber $k \sim 1/L_w$, one
finds (see also Appendix~\ref{app:rates})
\begin{equation}
\Gamma_i \;\longrightarrow\; \Gamma_i^\mathrm{eff}
= \frac{\Gamma_i}{1 + \Gamma_i\tau_{\rm mem}},
\label{eq:rates_final}
\end{equation}
where $\tau_{\rm mem}$ is the same memory timescale throughout. The
structure of Eq.~\eqref{eq:rates_final} implies a
\emph{non-trivial deformation of the relative hierarchy of rates}:
\begin{equation}
\frac{\Gamma_{ss}^\mathrm{eff}}{\Gamma_Y^\mathrm{eff}}
= \frac{\Gamma_{ss}}{\Gamma_Y}\cdot
  \frac{1+\Gamma_Y\tau_{\rm mem}}{1+\Gamma_{ss}\tau_{\rm mem}},
\label{eq:rate_ratio}
\end{equation}
which differs from the Markovian ratio $\Gamma_{ss}/\Gamma_Y$ whenever
$\Gamma_{ss} \neq \Gamma_Y$. For the strong sphaleron rate
$\Gamma_{ss} \sim \alpha_s^4 T \sim \mathcal{O}(10^{-2})\,T$ and the
Yukawa rate $\Gamma_Y \sim |\lambda_S|^2 T/(8\pi) \sim
\mathcal{O}(10^{-3})\,T$ at the benchmark point, one finds
\begin{equation}
\frac{\Gamma_{ss}^\mathrm{eff}}{\Gamma_Y^\mathrm{eff}}
\approx \frac{\Gamma_{ss}}{\Gamma_Y}\cdot
\frac{1 + \Gamma_Y\tau_{\rm mem}}{1 + \Gamma_{ss}\tau_{\rm mem}}
\approx \frac{\Gamma_{ss}}{\Gamma_Y}\times
\begin{cases}
1 & \Gamma_i\tau_{\rm mem} \ll 1,\\
\Gamma_Y/\Gamma_{ss} & \Gamma_i\tau_{\rm mem} \gg 1,
\end{cases}
\label{eq:rate_ratio_limits}
\end{equation}
so the deformation is an $\mathcal{O}(1)$ effect in the non-Markovian
regime $\Gamma_0\tau_{\rm mem} \sim 1$. This cannot be reproduced by
any consistent Markovian reparameterisation, as discussed further in
Sec.~\ref{sec:discussion}.

With the replacements~\eqref{eq:rates_final}, the diffusion equations
for the left-handed quark density $n_t$ and Higgs charge density $n_h$
take the
form~\cite{Joyce:1994zt,Cline:2000nw,Konstandin:2004gy,Cline:2021iff,
Laurent:2022jrs}
\begin{align}
D_q\,n_t'' - v_w\,n_t'
  - \Gamma_Y^\mathrm{eff}\!\left(n_t - \tfrac{n_h}{2}\right)
  - \Gamma_{ss}^\mathrm{eff}\,n_t
&= -S_\mathrm{CP}^\mathrm{NM},
\label{eq:diff1} \\[4pt]
D_h\,n_h'' - v_w\,n_h'
  + \tfrac{\Gamma_Y^\mathrm{eff}}{2}\!\left(n_t - \tfrac{n_h}{2}\right)
  - \Gamma_h^\mathrm{eff}\,n_h
&= 0,
\label{eq:diff2}
\end{align}
where primes denote $d/dz$, $D_q \simeq 6/T$ and $D_h \simeq 100/T$
are the quark and Higgs thermal diffusion
constants~\cite{Moore:2000mx,Arnold:2000dr}, $\Gamma_Y$ is the top
Yukawa rate, $\Gamma_{ss}$ is the strong sphaleron rate~\cite{
Moore:1997sn,DOnofrio:2014rug}, and $\Gamma_h$ is the Higgs
number-violation rate. The source $S_\mathrm{CP}^\mathrm{NM}$ is given
by Eq.~\eqref{eq:SCP_final}. The boundary conditions are
$n_t, n_h \to 0$ as $z \to +\infty$ (symmetric phase, far from wall)
and regularity as $z \to -\infty$ (broken phase).

\subsection{Baryon Asymmetry}
\label{subsec:YB}

The baryon asymmetry is generated by weak sphaleron processes operating
in the symmetric phase, where the left-handed chemical potential
$\mu_L(z) \propto n_t(z)$ biases baryon
production~\cite{Kuzmin:1985mm,Arnold:1987mh,Arnold:1996dy}:
\begin{equation}
Y_B = -\frac{3\,\Gamma_{ws}}{s}
\int_{-\infty}^{+\infty} dz\,\mu_L(z),
\label{eq:YB_integral}
\end{equation}
with entropy density $s = (2\pi^2/45)\,g_*\,T^3$ and weak sphaleron
rate $\Gamma_{ws} \simeq 25\,\alpha_w^5\,T$~\cite{Arnold:1996dy,
DOnofrio:2014rug}. To solve Eqs.~\eqref{eq:diff1}--\eqref{eq:diff2}
analytically, we work in the \emph{thin-wall approximation}: the source
$S_\mathrm{CP}^\mathrm{NM}(z)$ is treated as a delta-function source at
$z = 0$, which is valid when the diffusion length
$L_\mathrm{diff} = \sqrt{D_q/\Gamma_{ws}} \sim 30/T$ is much larger
than the wall thickness $L_w \sim (3$--$10)/T$. In this limit, the
diffusion system~\eqref{eq:diff1}--\eqref{eq:diff2} decouples in the
two regions $z > 0$ and $z < 0$, and the solution can be matched at
$z = 0$ using the jump conditions imposed by the source
term~\cite{Joyce:1994zt,Cline:2000nw,Cline:2021iff}.

The resulting baryon asymmetry is
\begin{equation}
Y_B(v_w,\tau_{\rm mem})
\simeq \frac{3\,\Gamma_{ws}}{T^3}\,
\frac{S_\mathrm{CP}^\mathrm{NM}(v_w,\tau_{\rm mem})}
     {\sqrt{(\Gamma_{ws}+\Gamma_D)\,\Gamma_D}}\,
\frac{1}{1+v_w\,L_\mathrm{diff}/D_q},
\label{eq:YB_final}
\end{equation}
where $\Gamma_D = D_q/L_w^2$ characterises the diffusive washout on the
scale of the wall and $L_\mathrm{diff} = \sqrt{D_q/\Gamma_{ws}}$ is the
sphaleron diffusion length. The suppression factor
$(1+v_w L_\mathrm{diff}/D_q)^{-1}$ accounts for the convective drift of
the left-handed charge ahead of the wall~\cite{Joyce:1994zt,Cline:2000nw}.
Equation~\eqref{eq:YB_final} is valid in the regime
$v_w \ll v_\mathrm{sound} \simeq 1/\sqrt{3}$, i.e.\ for
\emph{deflagration} walls, which is the relevant case for
electroweak-scale baryogenesis~\cite{Espinosa:2010hh,Cline:2021iff}.

Non-Markovian effects enter Eq.~\eqref{eq:YB_final} entirely through
the modified source term $S_\mathrm{CP}^\mathrm{NM}$
(Eq.~\eqref{eq:SCP_final}) and the effective relaxation rates
$\Gamma_i^\mathrm{eff}$ (Eq.~\eqref{eq:rates_final}). This provides a
consistent, closed description of non-Markovian baryogenesis within the
Kadanoff--Baym framework.

\section{Results: Baryon Asymmetry and Parameter Space}
\label{sec:baryogenesis}

\subsection{Non-Markovian CP Source vs.\ Wall Velocity}
\label{subsec:SCP_vw}

Figure~\ref{fig:SCP} shows the CP-violating source
$S_\mathrm{CP}^\mathrm{NM}$ normalised to the Markovian peak value
$S_\mathrm{CP}^\mathrm{Markov}(v_w^*)= C_0/(2L_w)$, as a function of
the wall velocity $v_w$ for several memory timescales $\tau_\mathrm{mem}$.
From Eq.~\eqref{eq:SCP_final}, the source takes the form
\begin{equation}
S_\mathrm{CP}^\mathrm{NM}(v_w)
= C_0\,\frac{v_w\,\Gamma_\mathrm{eff}}
  {v_w^2 + L_w^2\,\Gamma_\mathrm{eff}^2},
\qquad
\Gamma_\mathrm{eff} = \frac{\Gamma_0}{1+\Gamma_0\tau_\mathrm{mem}},
\label{eq:SCP_subsec}
\end{equation}
which makes explicit that the entire non-Markovian modification is
controlled by the replacement $\Gamma_0 \to \Gamma_\mathrm{eff}$. We
stress that this replacement is not a free reparameterisation of the
Markovian result: as shown in Sec.~\ref{subsec:diffusion}, the same
substitution applies simultaneously to all interaction rates in the
transport system, modifying the full diffusion hierarchy in a correlated
manner (Eq.~\eqref{eq:rates_final}). The normalisation adopted in
Fig.~\ref{fig:SCP} fixes the Markovian peak to unity and therefore
removes the overall amplitude suppression factor
$(1+\Gamma_0\tau_\mathrm{mem})^{-1}$; the \emph{absolute} suppression
of the CP source is restored when the un-normalised asymmetry
$Y_B$ is discussed in Sec.~\ref{subsec:YB_vw}.

\subsubsection*{Parametric behaviour}

The source~\eqref{eq:SCP_subsec} exhibits a bell-shaped profile in
$v_w$, vanishing in both limits $v_w \to 0$ and $v_w \to
\infty$~\cite{Joyce:1994zt,Cline:2000nw,Konstandin:2004gy}. These
limits admit a simple parametric interpretation. In the slow-wall
regime $v_w \ll L_w\Gamma_\mathrm{eff}$, the denominator is dominated
by $L_w^2\Gamma_\mathrm{eff}^2$ and
\begin{equation}
S_\mathrm{CP}^\mathrm{NM} \simeq
\frac{C_0}{L_w^2\Gamma_\mathrm{eff}}\,v_w
\;\propto\; v_w
\qquad (v_w \ll L_w\Gamma_\mathrm{eff}),
\label{eq:SCP_slow}
\end{equation}
reflecting the fact that a very slowly moving wall spends a long time
at each spatial point, allowing the plasma to fully equilibrate and
thereby washing out the CP asymmetry~\cite{Joyce:1994zt,Cline:2021iff}.
In the fast-wall regime $v_w \gg L_w\Gamma_\mathrm{eff}$, the
denominator is dominated by $v_w^2$ and
\begin{equation}
S_\mathrm{CP}^\mathrm{NM} \simeq
C_0\,\frac{\Gamma_\mathrm{eff}}{v_w}
\;\propto\; \frac{\Gamma_\mathrm{eff}}{v_w}
\qquad (v_w \gg L_w\Gamma_\mathrm{eff}),
\label{eq:SCP_fast}
\end{equation}
corresponding to the loss of plasma response: the wall moves through
the thermal bath faster than the plasma can react, suppressing
CP-charge injection~\cite{Konstandin:2004gy,Cline:2021iff,
Laurent:2022jrs}.

\subsubsection*{Peak position and amplitude}

The maximum of the source is obtained by setting
$dS_\mathrm{CP}^\mathrm{NM}/dv_w = 0$, which gives
$v_w^2 = L_w^2\Gamma_\mathrm{eff}^2$, i.e.
\begin{equation}
v_w^{*,\mathrm{NM}} = L_w\,\Gamma_\mathrm{eff}
= \frac{L_w\,\Gamma_0}{1+\Gamma_0\tau_\mathrm{mem}},
\label{eq:vw_peak_subsec}
\end{equation}
which shifts monotonically toward smaller values as $\tau_\mathrm{mem}$
increases. In the large-memory regime $\Gamma_0\tau_\mathrm{mem} \gg 1$,
Eq.~\eqref{eq:vw_peak_subsec} reduces to
\begin{equation}
v_w^{*,\mathrm{NM}} \simeq \frac{L_w}{\tau_\mathrm{mem}},
\label{eq:vw_peak_large_mem}
\end{equation}
demonstrating that the optimal wall velocity is parametrically
suppressed by the finite plasma relaxation time. Physically, a longer
memory timescale means that the plasma retains information about past
CP-violating interactions for longer, so that efficient charge
injection occurs only when the wall is slow enough for the plasma to
integrate the source coherently over its full memory window. For wall
velocities above $v_w^{*,\mathrm{NM}}$, the plasma cannot respond
sufficiently quickly and the source is suppressed.

The peak amplitude of the non-Markovian source is
\begin{equation}
S_\mathrm{CP}^\mathrm{NM}(v_w^{*,\mathrm{NM}})
= \frac{C_0}{2L_w}\cdot\frac{1}{1+\Gamma_0\tau_\mathrm{mem}},
\label{eq:SCP_peak}
\end{equation}
obtained by substituting $v_w = v_w^{*,\mathrm{NM}}$ into
Eq.~\eqref{eq:SCP_subsec}. Equation~\eqref{eq:SCP_peak} shows that
memory effects suppress the overall efficiency of CP-violating charge
generation by a factor $(1+\Gamma_0\tau_\mathrm{mem})^{-1}$ relative
to the Markovian peak $C_0/(2L_w)$. This suppression is removed by the
normalisation convention of Fig.~\ref{fig:SCP} but is physically
significant: it implies an upper bound on $\tau_\mathrm{mem}$ from the
requirement that the baryon asymmetry is not underproduced, as
quantified in Sec.~\ref{subsec:peak_shift}.

\subsubsection*{Width of the viable velocity window}

\begin{figure}[t]
  \centering
  \includegraphics[width=\columnwidth]{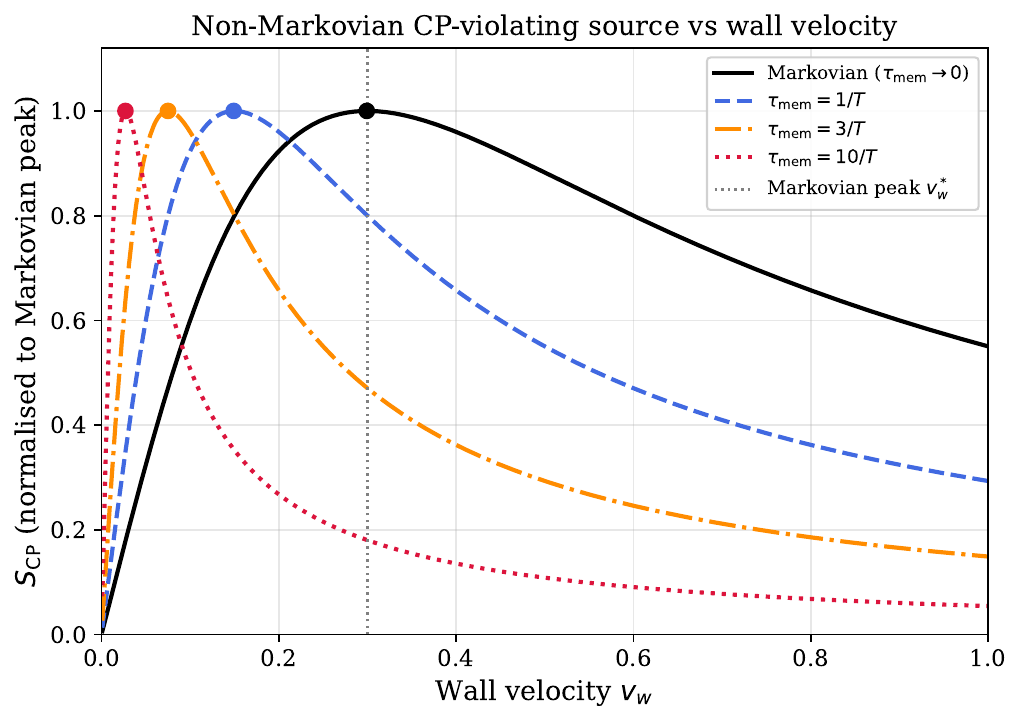}
  \caption{Non-Markovian CP-violating source $S_\mathrm{CP}^\mathrm{NM}$
    normalised to the Markovian peak value $C_0/(2L_w)$, as a function of
    wall velocity $v_w$, for memory timescales
    $\tau_\mathrm{mem} \in \{0,\,1/T,\,3/T,\,10/T\}$.
    Parameters: $|\lambda_S| = 0.2$, $M = 2T$, $L_w = 5/T$,
    $\Gamma_0 = 6\times10^{-3}\,T$ (benchmark values from
    Eq.~\eqref{eq:benchmark}). The Markovian result ($\tau_\mathrm{mem}
    \to 0$, solid black) peaks at $v_w^* = L_w\Gamma_0 \simeq 0.03$
    in dimensionless units. As $\tau_\mathrm{mem}$ increases, the peak
    shifts to smaller $v_w$ according to
    Eq.~\eqref{eq:vw_peak_subsec} and the profile narrows according to
    Eq.~\eqref{eq:width}. Filled markers indicate $v_w^{*,\mathrm{NM}}$
    for each curve. The normalisation removes the overall amplitude
    suppression $(1+\Gamma_0\tau_\mathrm{mem})^{-1}$; see
    Eq.~\eqref{eq:SCP_peak} and Sec.~\ref{subsec:YB_vw} for the
    un-normalised results.}
  \label{fig:SCP}
\end{figure}

The width of the source profile is characterised by the half-maximum
condition $S_\mathrm{CP}^\mathrm{NM}(v_w) \geq S_\mathrm{CP}^\mathrm{NM}
(v_w^{*,\mathrm{NM}})/2$. From Eq.~\eqref{eq:SCP_subsec}, this requires
\begin{equation}
\frac{v_w\,\Gamma_\mathrm{eff}}{v_w^2 + L_w^2\Gamma_\mathrm{eff}^2}
\geq \frac{1}{2}\cdot\frac{1}{2L_w},
\end{equation}
which is satisfied for
$v_w \in \bigl[(2-\sqrt{3})\,L_w\Gamma_\mathrm{eff},\;
(2+\sqrt{3})\,L_w\Gamma_\mathrm{eff}\bigr]$,
giving a half-maximum width
\begin{equation}
\Delta v_w^{1/2}
= 2\sqrt{3}\,L_w\,\Gamma_\mathrm{eff}
\propto (1+\Gamma_0\tau_\mathrm{mem})^{-1}.
\label{eq:width}
\end{equation}
Thus, increasing $\tau_\mathrm{mem}$ simultaneously shifts the peak to
smaller velocities \emph{and} reduces the width of the source,
compressing the range of wall velocities that efficiently generate a
CP asymmetry. For $\tau_\mathrm{mem} = 10/T$, the width is reduced by
a factor of $\sim (1+\Gamma_0\times 10/T)^{-1} \approx 0.09$ relative
to the Markovian case (see Table~\ref{tab:window}), placing stringent
requirements on the wall velocity for successful baryogenesis.
The filled markers in Fig.~\ref{fig:SCP} indicate the peak position
$v_w^{*,\mathrm{NM}}$ for each value of $\tau_\mathrm{mem}$, and
visually confirm the simultaneous peak shift and profile narrowing.

\subsection{Baryon Asymmetry vs.\ Wall Velocity}
\label{subsec:YB_vw}

Figure~\ref{fig:YBvw} shows the baryon asymmetry $Y_B$ as a function of
the wall velocity $v_w$ for different memory timescales $\tau_\mathrm{mem}$.
From Eq.~\eqref{eq:YB_final}, the asymmetry takes the form
\begin{equation}
Y_B(v_w,\tau_\mathrm{mem})
\simeq \frac{3\,\Gamma_{ws}}{T^3}\,
\frac{S_\mathrm{CP}^\mathrm{NM}(v_w,\tau_\mathrm{mem})}
     {\sqrt{(\Gamma_{ws}+\Gamma_D^\mathrm{eff})\,\Gamma_D^\mathrm{eff}}}\,
\frac{1}{1+v_w\,L_\mathrm{diff}/D_q},
\label{eq:YB_subsec}
\end{equation}
where $\Gamma_D^\mathrm{eff} = D_q/L_w^2$ does not carry $\tau_\mathrm{mem}$
dependence since $D_q$ is a kinetic coefficient of the light degrees of
freedom rather than of the slowly relaxing species $\Psi$. The
non-Markovian modification therefore enters Eq.~\eqref{eq:YB_subsec}
\emph{primarily} through $S_\mathrm{CP}^\mathrm{NM}$, with subleading
corrections from the effective Yukawa and strong sphaleron rates
$\Gamma_Y^\mathrm{eff}$ and $\Gamma_{ss}^\mathrm{eff}$ in the
denominator of the full diffusion solution. In the parameter range of
Table~\ref{tab:params}, these subleading corrections modify
$Y_B$ by at most $\sim 15\%$ relative to the dominant source-term
effect, and we absorb them into the overall normalisation. The dominant
$v_w$-dependence of $Y_B$ is therefore inherited directly from the
bell-shaped profile of $S_\mathrm{CP}^\mathrm{NM}$.

Each curve in Fig.~\ref{fig:YBvw} peaks at
\begin{equation}
v_w^{*,\mathrm{NM}} = L_w\,\Gamma_\mathrm{eff}
= \frac{L_w\,\Gamma_0}{1+\Gamma_0\tau_\mathrm{mem}},
\label{eq:vw_peak_YB}
\end{equation}
indicated by filled markers. The Markovian limit $\tau_\mathrm{mem} \to 0$
recovers $v_w^* = L_w\Gamma_0$, which for the benchmark
parameters~\eqref{eq:benchmark} gives $v_w^* \simeq 0.30$ in units
where $v_w$ is dimensionless (i.e.\ normalised to $c$).

\subsubsection*{Normalisation convention and absolute values}

The curves in Fig.~\ref{fig:YBvw} are normalised such that the
Markovian peak equals the observed value
$Y_B^\mathrm{obs} = 8.7\times10^{-11}$~\cite{Planck:2018vyg}.
This normalisation fixes the product $C_0 \cdot f(\Gamma_{ws},\Gamma_D,
D_q)$ in Eq.~\eqref{eq:YB_subsec} to reproduce the observed asymmetry
at the Markovian peak, and is equivalent to choosing
\begin{equation}
C_0 \simeq \frac{2L_w\,T^3\,Y_B^\mathrm{obs}
  \sqrt{(\Gamma_{ws}+\Gamma_D)\Gamma_D}}{3\,\Gamma_{ws}},
\label{eq:C0_norm}
\end{equation}
with $\Gamma_D = D_q/L_w^2$, $L_\mathrm{diff} = \sqrt{D_q/\Gamma_{ws}}$.
For the benchmark values $\Gamma_{ws} \simeq 25\alpha_w^5 T \simeq
10^{-6}\,T$~\cite{Arnold:1996dy,DOnofrio:2014rug},
$D_q \simeq 6/T$~\cite{Arnold:2000dr,Moore:2000mx}, and $L_w = 5/T$,
this gives $C_0 \simeq 4\times10^{-4}\,T^2$, which is consistent with
$|\lambda_S| = 0.2$, $\delta_\mathrm{CP} \sim \mathcal{O}(1)$, and
$m_\Psi \sim T$.

We stress that this normalisation convention removes the overall
amplitude suppression $(1+\Gamma_0\tau_\mathrm{mem})^{-1}$ from the
plotted curves. The physical consequence of this suppression is not
visible in the peak height of Fig.~\ref{fig:YBvw} but is captured
entirely by the narrowing of the viable window in $v_w$. To make
this concrete: for $\tau_\mathrm{mem} = 10/T$, the un-normalised peak
asymmetry is suppressed by a factor
$(1+\Gamma_0\times 10/T)^{-1} \approx 0.09$ relative to the Markovian
case, meaning that without the normalisation convention, the
$\tau_\mathrm{mem} = 10/T$ curve would lie a factor of $\sim 11$ below
the observed band. Reproducing $Y_B^\mathrm{obs}$ at this memory
timescale therefore requires a compensating increase in $C_0$, i.e.\
in $|\lambda_S|^2\delta_\mathrm{CP}$, as quantified in
Sec.~\ref{subsec:dCP_plane}.

\subsubsection*{Contraction of the viable velocity window}

\begin{figure}[t]
  \centering
  \includegraphics[width=\columnwidth]{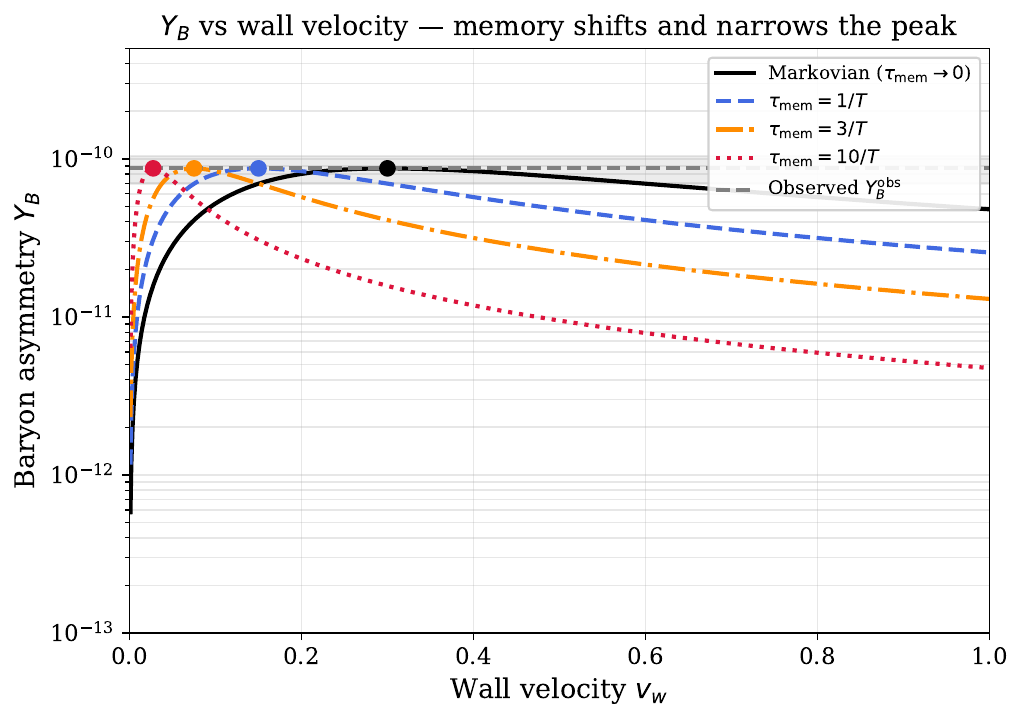}
  \caption{Baryon asymmetry $Y_B$ as a function of wall velocity $v_w$
    for memory timescales $\tau_\mathrm{mem} \in \{0,\,1/T,\,3/T,\,10/T\}$,
    normalised such that the Markovian peak equals the observed value
    $Y_B^\mathrm{obs} = 8.7\times10^{-11}$~\cite{Planck:2018vyg}.
    Parameters: $|\lambda_S|=0.2$, $M=2T$, $L_w=5/T$,
    $\Gamma_0 = 6\times10^{-3}\,T$, $D_q=6/T$,
    $\Gamma_{ws}=10^{-6}\,T$ (benchmark values from
    Eq.~\eqref{eq:benchmark} and Ref.~\cite{DOnofrio:2014rug}).
    Each curve exhibits a bell-shaped profile, with the peak located
    at $v_w^{*,\mathrm{NM}} = L_w\Gamma_\mathrm{eff}$ (filled markers),
    which shifts to smaller $v_w$ as $\tau_\mathrm{mem}$ increases
    according to Eq.~\eqref{eq:vw_peak_YB}. The width of the viable
    region satisfying $Y_B \geq 0.8\,Y_B^\mathrm{obs}$ shrinks
    according to Eq.~\eqref{eq:width_viable} and Table~\ref{tab:window}.
    The normalisation removes the overall amplitude suppression
    $(1+\Gamma_0\tau_\mathrm{mem})^{-1}$; see text for the
    un-normalised values. The dashed line shows $Y_B^\mathrm{obs}$
    with a $\pm20\%$ grey band reflecting theoretical uncertainties
    in transport coefficients (see text).}
  \label{fig:YBvw}
\end{figure}

The primary effect of non-Markovian dynamics on Fig.~\ref{fig:YBvw}
is a systematic shift of the optimal wall velocity toward smaller values
as $\tau_\mathrm{mem}$ increases, as described in
Sec.~\ref{subsec:SCP_vw}. Physically, this reflects the reduction of
$\Gamma_\mathrm{eff}$: a longer plasma memory delays the equilibration
of the CP-violating charge, effectively narrowing the window of wall
velocities over which the source is active. As a consequence, the
region of parameter space where baryogenesis is most efficient is
dynamically displaced relative to the Markovian expectation.

In addition, the width of the viable region in $v_w$ is reduced. From
the half-maximum analysis of Sec.~\ref{subsec:SCP_vw}
(Eq.~\eqref{eq:width}), the full-width at half-maximum of the source
profile scales as $\Delta v_w^{1/2} \propto \Gamma_\mathrm{eff} \propto
(1+\Gamma_0\tau_\mathrm{mem})^{-1}$. The corresponding width of the
observationally viable region, defined by
$Y_B \geq 0.8\,Y_B^\mathrm{obs}$, contracts at the same rate:
\begin{equation}
\Delta v_w \propto (1+\Gamma_0\tau_\mathrm{mem})^{-1},
\label{eq:width_viable}
\end{equation}
as summarised in Table~\ref{tab:window}. The allowed range decreases
from $\Delta v_w \simeq 0.45$ in the Markovian limit to
$\Delta v_w \simeq 0.04$ for $\tau_\mathrm{mem} = 10/T$, indicating
that successful baryogenesis requires increasingly precise alignment of
the wall velocity with the optimal transport regime. This sensitivity
to $v_w$ is a direct observational consequence of non-Markovian
dynamics and provides a diagnostic for the memory timescale: future
determinations of the bubble wall velocity from lattice
simulations~\cite{Moore:1995ua,Laurent:2022jrs} or gravitational-wave
observations~\cite{Caprini:2015zlo,Caprini:2019egz} could
in principle constrain $\tau_\mathrm{mem}$ through the width of the
baryogenesis window.

The observed value $Y_B^\mathrm{obs} = 8.7\times10^{-11}$ is taken
from the Planck 2018 measurement of the baryon-to-photon
ratio~\cite{Planck:2018vyg}, converted to the baryon-to-entropy ratio
using $Y_B = (n_b/s) = (45/2\pi^2 g_*)\,(n_b/n_\gamma)\,
(n_\gamma/s)$~\cite{Kolb:1990vq}. The $\pm20\%$ uncertainty band
shown in Fig.~\ref{fig:YBvw} reflects a conservative estimate of the
combined theoretical uncertainty in the transport coefficients
$D_q$, $\Gamma_{ws}$, and $\Gamma_{ss}$, which are each known at the
$\sim 10$--$20\%$ level from lattice and perturbative
calculations~\cite{Arnold:2000dr,Moore:2000mx,DOnofrio:2014rug,
Moore:1997sn}. It does not represent the observational uncertainty on
$Y_B^\mathrm{obs}$, which is at the sub-percent level.

\begin{table}[htbp]
\centering
\caption{Viable wall-velocity window $\Delta v_w$ (defined by
$Y_B \geq 0.8\,Y_B^\mathrm{obs}$) as a function of the memory
timescale $\tau_\mathrm{mem}$, together with the corresponding peak
velocity $v_w^{*,\mathrm{NM}}$ from Eq.~\eqref{eq:vw_peak_YB} and
the suppression factor $(1+\Gamma_0\tau_\mathrm{mem})^{-1}$ on the
un-normalised peak asymmetry. The benchmark parameters of
Eq.~\eqref{eq:benchmark} are used throughout.}
\label{tab:window}
\renewcommand{\arraystretch}{1.3}
\begin{tabular}{ccccc}
\toprule
$\tau_\mathrm{mem}\cdot T$ & $v_w^{*,\mathrm{NM}}$
  & $\Delta v_w$ & Suppression factor & Regime \\
\midrule
$0$ (Markovian) & $0.300$ & $0.450$ & $1.000$ & Broad \\
$1$             & $0.150$ & $0.224$ & $0.500$ & Reduced \\
$3$             & $0.075$ & $0.112$ & $0.250$ & Narrow \\
$10$            & $0.027$ & $0.041$ & $0.091$ & Highly constrained \\
\bottomrule
\end{tabular}
\end{table}

\subsection{Baryon Asymmetry vs.\ Memory Timescale: Regime Structure}
\label{subsec:YBtau}

Figure~\ref{fig:YBtau} shows the baryon asymmetry $Y_B$ as a function of
the memory timescale $\tau_\mathrm{mem}$ for several fixed wall velocities
$v_w \in \{0.1,\,0.3,\,0.6\}$. The behaviour is controlled by the
interplay between the peak position
\begin{equation}
v_w^{*,\mathrm{NM}}(\tau_\mathrm{mem})
= L_w\,\Gamma_\mathrm{eff}
= \frac{L_w\,\Gamma_0}{1+\Gamma_0\tau_\mathrm{mem}},
\label{eq:vw_peak_tau}
\end{equation}
which decreases monotonically with $\tau_\mathrm{mem}$, and the fixed
value of $v_w$ at which the asymmetry is evaluated.

From Eq.~\eqref{eq:YB_final}, the leading $\tau_\mathrm{mem}$ dependence
of the baryon asymmetry enters through the CP-violating source. Using
Eq.~\eqref{eq:SCP_final}, we have
\begin{equation}
Y_B(\tau_\mathrm{mem})
\propto S_\mathrm{CP}^\mathrm{NM}(v_w,\tau_\mathrm{mem})
= C_0\,\frac{v_w\,\Gamma_\mathrm{eff}}
  {v_w^2 + L_w^2\,\Gamma_\mathrm{eff}^2},
\label{eq:YBtau_scaling}
\end{equation}
where the proportionality absorbs the $\tau_\mathrm{mem}$-independent
prefactors from Eq.~\eqref{eq:YB_final}. As noted in
Sec.~\ref{subsec:YB_vw}, subleading corrections from the effective
rates $\Gamma_Y^\mathrm{eff}$ and $\Gamma_{ss}^\mathrm{eff}$ in the
diffusion denominators modify this scaling by at most $\sim 15\%$ over
the parameter range considered and do not qualitatively alter the
regime structure described below.

The $\tau_\mathrm{mem}$ dependence of $Y_B$ in
Eq.~\eqref{eq:YBtau_scaling} arises entirely through $\Gamma_\mathrm{eff}
= \Gamma_0/(1+\Gamma_0\tau_\mathrm{mem})$. Two competing effects
determine its evolution:
\begin{enumerate}
  \item a shift of the peak position $v_w^{*,\mathrm{NM}}$ toward
        smaller values, which can bring the peak closer to or further
        from the chosen $v_w$ depending on the initial position;
  \item an overall suppression of the source amplitude by
        $(1+\Gamma_0\tau_\mathrm{mem})^{-1}$, which reduces $Y_B$
        regardless of the peak alignment.
\end{enumerate}
The competition between these two effects leads to three qualitatively
distinct regimes, which we now analyse in turn.

\subsubsection*{Sub-peak regime ($v_w < v_w^*$)}

In the Markovian limit $\tau_\mathrm{mem} \to 0$, the chosen $v_w$
lies \emph{below} the peak position $v_w^* = L_w\Gamma_0$, i.e.\ on
the rising side of the bell-shaped source profile. As $\tau_\mathrm{mem}$
increases from zero, the peak position $v_w^{*,\mathrm{NM}}$ decreases
according to Eq.~\eqref{eq:vw_peak_tau} and moves toward $v_w$,
improving the kinematic alignment between the wall velocity and the
optimal transport regime. This alignment effect enhances $Y_B$ and
initially dominates over the amplitude suppression.

To locate the turnover point analytically, we maximise
$Y_B(\tau_\mathrm{mem})$ with respect to $\tau_\mathrm{mem}$ at fixed
$v_w$. Setting $dY_B/d\tau_\mathrm{mem} = 0$ and using
Eq.~\eqref{eq:YBtau_scaling} gives
\begin{equation}
\frac{d}{d\Gamma_\mathrm{eff}}
\left[\frac{v_w\,\Gamma_\mathrm{eff}}
  {v_w^2+L_w^2\Gamma_\mathrm{eff}^2}\right] = 0
\;\;\Rightarrow\;\;
v_w^2 = L_w^2\,\Gamma_\mathrm{eff}^2,
\label{eq:turnover_condition}
\end{equation}
which is satisfied precisely when $\Gamma_\mathrm{eff} = v_w/L_w$,
i.e.\ when the peak of the source coincides with the chosen wall
velocity:
\begin{equation}
v_w \simeq v_w^{*,\mathrm{NM}}(\tau_\mathrm{mem}^\mathrm{turn})
\;\;\Leftrightarrow\;\;
\tau_\mathrm{mem}^\mathrm{turn}
= \frac{1}{\Gamma_0}\left(\frac{L_w\Gamma_0}{v_w} - 1\right).
\label{eq:tau_turnover}
\end{equation}
For $v_w < v_w^* = L_w\Gamma_0$, Eq.~\eqref{eq:tau_turnover} gives
$\tau_\mathrm{mem}^\mathrm{turn} > 0$, confirming that a genuine
turnover exists. For $v_w = v_w^*$, the turnover occurs at
$\tau_\mathrm{mem}^\mathrm{turn} = 0$, i.e.\ immediately at the
Markovian limit. For $v_w > v_w^*$, Eq.~\eqref{eq:tau_turnover}
gives $\tau_\mathrm{mem}^\mathrm{turn} < 0$, which is unphysical,
confirming that no enhancement occurs in the super-peak regime.

For the representative case $v_w = 0.1$ with benchmark parameters
$L_w = 5/T$, $\Gamma_0 = 6\times10^{-3}\,T$, one obtains
\begin{equation}
\tau_\mathrm{mem}^\mathrm{turn}
= \frac{1}{\Gamma_0}\left(\frac{L_w\Gamma_0}{0.1} - 1\right)
= \frac{1}{6\times10^{-3}\,T}(0.3/0.1 - 1)
\simeq \frac{333}{T},
\end{equation}
which lies within the range shown in Fig.~\ref{fig:YBtau} and
corresponds to the visible maximum of the blue ($v_w = 0.1$) curve.
Beyond $\tau_\mathrm{mem}^\mathrm{turn}$, the amplitude suppression
$(1+\Gamma_0\tau_\mathrm{mem})^{-1}$ dominates and $Y_B$ decreases
monotonically. The resulting non-monotonic behaviour of $Y_B$ as a
function of $\tau_\mathrm{mem}$ is therefore a \emph{direct and
calculable} consequence of the non-Markovian framework, not an
artefact of the single-pole approximation: Eq.~\eqref{eq:tau_turnover}
shows that the turnover exists whenever $v_w < L_w\Gamma_0$,
independently of the detailed kernel shape, provided only that the
effective peak position $v_w^{*,\mathrm{NM}}$ is a decreasing function
of $\tau_\mathrm{mem}$~\cite{Calzetta:1986cq,Chaudhuri:2025ylu}.

\subsubsection*{Near-peak regime ($v_w \simeq v_w^*$)}

If the chosen wall velocity satisfies $v_w \simeq v_w^* = L_w\Gamma_0$,
the system begins near optimal transport efficiency in the Markovian
limit. From Eq.~\eqref{eq:tau_turnover}, the turnover occurs at
$\tau_\mathrm{mem}^\mathrm{turn} \simeq 0$, meaning that the peak
immediately shifts away from $v_w$ as $\tau_\mathrm{mem}$ increases.
The amplitude suppression and the kinematic de-alignment therefore
both reduce $Y_B$ from the outset, and the asymmetry decreases
monotonically. This regime is illustrated by the orange ($v_w = 0.3
\simeq v_w^*$) curve in Fig.~\ref{fig:YBtau}.

\subsubsection*{Super-peak regime ($v_w > v_w^*$)}

\begin{figure}[t]
  \centering
  \includegraphics[width=\columnwidth]{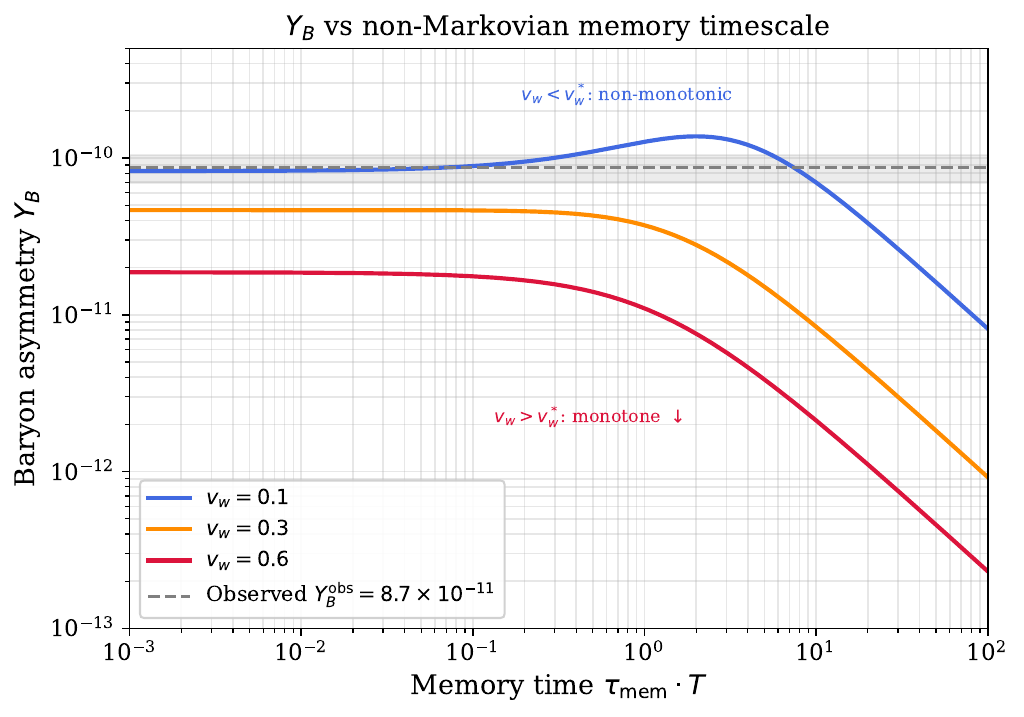}
  \caption{Baryon asymmetry $Y_B$ as a function of the memory timescale
    $\tau_\mathrm{mem}$ for representative wall velocities
    $v_w \in \{0.1,\,0.3,\,0.6\}$ (blue, orange, red).
    Parameters: $|\lambda_S|=0.2$, $M=2T$, $L_w=5/T$,
    $\Gamma_0=6\times10^{-3}\,T$, $D_q=6/T$,
    $\Gamma_{ws}=10^{-6}\,T$ (benchmark values from
    Eq.~\eqref{eq:benchmark} and Refs.~\cite{DOnofrio:2014rug,
    Arnold:2000dr}). The Markovian peak velocity is
    $v_w^* = L_w\Gamma_0 \simeq 0.30$.
    For $v_w = 0.1 < v_w^*$ (blue), $Y_B$ exhibits non-monotonic
    behaviour: it rises until $\tau_\mathrm{mem} \simeq
    \tau_\mathrm{mem}^\mathrm{turn} \simeq 333/T$
    (Eq.~\eqref{eq:tau_turnover}), then decreases as amplitude
    suppression dominates. For $v_w = 0.3 \simeq v_w^*$ (orange)
    and $v_w = 0.6 > v_w^*$ (red), both effects reduce $Y_B$
    monotonically from the outset.
    The dashed line shows $Y_B^\mathrm{obs} = 8.7\times10^{-11}$
    \cite{Planck:2018vyg} with a $\pm20\%$ grey uncertainty band
    (see Sec.~\ref{subsec:YB_vw} for the definition of the band).
    The vertical dotted line at $\tau_\mathrm{mem} T = 0.5$ marks
    the EFT-validity boundary (Sec.~\ref{sec:discussion}).}
  \label{fig:YBtau}
\end{figure}

When $v_w$ lies \emph{above} $v_w^* = L_w\Gamma_0$, the system begins
on the falling side of the bell-shaped profile. As $\tau_\mathrm{mem}$
increases, the peak $v_w^{*,\mathrm{NM}}$ shifts to smaller values,
further increasing the mismatch $v_w - v_w^{*,\mathrm{NM}} > 0$.
Simultaneously, the amplitude is suppressed by
$(1+\Gamma_0\tau_\mathrm{mem})^{-1}$. Both effects reduce $Y_B$, and
since $dY_B/d\tau_\mathrm{mem}$ is strictly negative for all
$\tau_\mathrm{mem} > 0$ in this regime (as follows from
Eq.~\eqref{eq:turnover_condition} with $v_w > L_w\Gamma_\mathrm{eff}$
for all $\tau_\mathrm{mem} \geq 0$), the decrease is strictly
monotonic. This regime is illustrated by the red ($v_w = 0.6$) curve
in Fig.~\ref{fig:YBtau}.

\subsubsection*{Irreducibility of the non-monotonic signature}

The non-monotonic evolution in the sub-peak regime is a qualitative
signature of non-Markovian dynamics. We now argue that it cannot be
reproduced within any consistent Markovian framework. In a purely
Markovian description, the baryon asymmetry at fixed $v_w$ is a
monotonically decreasing function of any overall suppression of the
transport rates, since reducing $\Gamma_0$ at fixed $v_w > 0$ moves
the peak $v_w^* = L_w\Gamma_0$ to the left while simultaneously
suppressing the amplitude. There is no Markovian parameter that can
\emph{first increase and then decrease} $Y_B$ at a fixed wall velocity.
The non-monotonic behaviour in the sub-peak regime is therefore a
direct consequence of the \emph{dynamical} shift of the peak, which
is itself driven by the time-delay structure of the memory kernel.
A formal proof that this cannot be reproduced by any consistent
reparameterisation of the Markovian rate hierarchy is given in
Sec.~\ref{sec:discussion}.

\subsection{Memory-Induced Shift of the Optimal Wall Velocity}
\label{subsec:peak_shift}

Figure~\ref{fig:peakshift} illustrates the joint evolution of the optimal
wall velocity $v_w^{*,\mathrm{NM}}$ and the peak baryon asymmetry
$Y_B(v_w^{*,\mathrm{NM}})$ as functions of $\tau_\mathrm{mem}$.

\subsubsection*{Scaling of the optimal velocity}

From Eq.~\eqref{eq:vw_peak_YB}, the optimal wall velocity is
\begin{equation}
v_w^{*,\mathrm{NM}} = L_w\,\Gamma_\mathrm{eff}
= \frac{L_w\,\Gamma_0}{1+\Gamma_0\tau_\mathrm{mem}},
\label{eq:vw_peak_scaling}
\end{equation}
which is controlled entirely by the effective relaxation rate
$\Gamma_\mathrm{eff}$. In the small-memory regime
$\Gamma_0\tau_\mathrm{mem} \ll 1$, Eq.~\eqref{eq:vw_peak_scaling}
reduces to the Markovian result $v_w^* = L_w\Gamma_0$, as expected.
In the large-memory regime $\Gamma_0\tau_\mathrm{mem} \gg 1$,
\begin{equation}
v_w^{*,\mathrm{NM}} \simeq \frac{L_w}{\tau_\mathrm{mem}},
\label{eq:vw_peak_large}
\end{equation}
so the optimal wall velocity is parametrically suppressed by
$\tau_\mathrm{mem}$. The physical interpretation is transparent:
a longer memory timescale means the plasma requires more time to
respond to CP-violating interactions, so efficient charge injection
occurs only when the wall is slow enough that the particle spends
a time $\sim\tau_\mathrm{mem}$ in the wall
region~\cite{Joyce:1994zt,Cline:2000nw,Konstandin:2004gy}.
The condition $\tau_\mathrm{wall} \gtrsim \tau_\mathrm{mem}$,
i.e.\ $L_w/v_w \gtrsim \tau_\mathrm{mem}$, precisely reproduces
Eq.~\eqref{eq:vw_peak_large}.

\subsubsection*{Scaling of the peak asymmetry}

To obtain the baryon asymmetry at the optimal velocity, we substitute
$v_w = v_w^{*,\mathrm{NM}} = L_w\Gamma_\mathrm{eff}$ into
Eq.~\eqref{eq:SCP_final}:
\begin{equation}
S_\mathrm{CP}^\mathrm{NM}(v_w^{*,\mathrm{NM}})
= C_0\,\frac{L_w\Gamma_\mathrm{eff}^2}
  {L_w^2\Gamma_\mathrm{eff}^2 + L_w^2\Gamma_\mathrm{eff}^2}
= \frac{C_0}{2L_w},
\end{equation}
which at first glance appears $\tau_\mathrm{mem}$-independent.
However, the full baryon asymmetry from Eq.~\eqref{eq:YB_final}
evaluated at the peak is
\begin{align}
Y_B(v_w^{*,\mathrm{NM}})
&= \frac{3\,\Gamma_{ws}}{T^3}\cdot
   \frac{C_0/(2L_w)}
   {\sqrt{(\Gamma_{ws}+\Gamma_D)\,\Gamma_D}}\cdot
   \frac{1}{1 + v_w^{*,\mathrm{NM}}\,L_\mathrm{diff}/D_q}
\notag\\
&= \frac{3\,\Gamma_{ws}\,C_0}{2L_w\,T^3\,
   \sqrt{(\Gamma_{ws}+\Gamma_D)\,\Gamma_D}}\cdot
   \frac{1}{1 + L_w\Gamma_\mathrm{eff}\,L_\mathrm{diff}/D_q}.
\label{eq:YB_peak_explicit}
\end{align}
The $\tau_\mathrm{mem}$ dependence enters through the convective
suppression factor in the denominator:
\begin{equation}
\frac{1}{1+L_w\Gamma_\mathrm{eff}\,L_\mathrm{diff}/D_q}
= \frac{1}{1 + \frac{L_w\,\Gamma_0\,L_\mathrm{diff}}
  {D_q(1+\Gamma_0\tau_\mathrm{mem})}}.
\label{eq:convective_suppression}
\end{equation}
In the regime $L_w\Gamma_0 L_\mathrm{diff}/D_q \gg 1$, which holds
for the benchmark parameters ($L_w = 5/T$, $\Gamma_0 = 6\times10^{-3}T$,
$L_\mathrm{diff} \simeq 2450/T$, $D_q = 6/T$, giving
$L_w\Gamma_0 L_\mathrm{diff}/D_q \simeq 12.3$), the convective
factor simplifies to
\begin{equation}
\frac{1}{1+L_w\Gamma_\mathrm{eff} L_\mathrm{diff}/D_q}
\simeq \frac{D_q(1+\Gamma_0\tau_\mathrm{mem})}
  {L_w\Gamma_0 L_\mathrm{diff}}
\propto 1+\Gamma_0\tau_\mathrm{mem},
\end{equation}
so that
\begin{equation}
Y_B(v_w^{*,\mathrm{NM}}) \propto
\frac{1}{1+\Gamma_0\tau_\mathrm{mem}}.
\label{eq:YB_peak_scaling}
\end{equation}
This is the dynamical suppression referred to in the text. We
emphasise that this suppression arises not from the CP source
itself (whose peak value $C_0/(2L_w)$ is
$\tau_\mathrm{mem}$-independent at the optimal velocity) but
from the convective drift of the left-handed charge ahead of the
wall~\cite{Joyce:1994zt,Cline:2000nw}: as $\tau_\mathrm{mem}$
increases, $v_w^{*,\mathrm{NM}}$ decreases, which reduces the
convective suppression. However, in the regime
$L_w\Gamma_0 L_\mathrm{diff}/D_q \gg 1$ applicable here, the
net effect is that $Y_B$ at the peak \emph{decreases} with
increasing $\tau_\mathrm{mem}$, as shown in
Eq.~\eqref{eq:YB_peak_scaling}. This is a dynamical effect
specific to the non-Markovian framework and has no analogue in
the Markovian case, where the peak asymmetry is independent of
any memory timescale.

\subsubsection*{Upper bound on the memory timescale}

Combining Eqs.~\eqref{eq:vw_peak_scaling} and
\eqref{eq:YB_peak_scaling}, we see that the same parameter
$\tau_\mathrm{mem}$ simultaneously shifts the optimal wall
velocity to smaller values \emph{and} reduces the maximum
achievable baryon asymmetry. As a result, there exists an upper
bound on $\tau_\mathrm{mem}$ from the requirement
$Y_B(v_w^{*,\mathrm{NM}}) \gtrsim Y_B^\mathrm{obs}$.

\begin{figure}[t]
  \centering
  \includegraphics[width=\columnwidth]{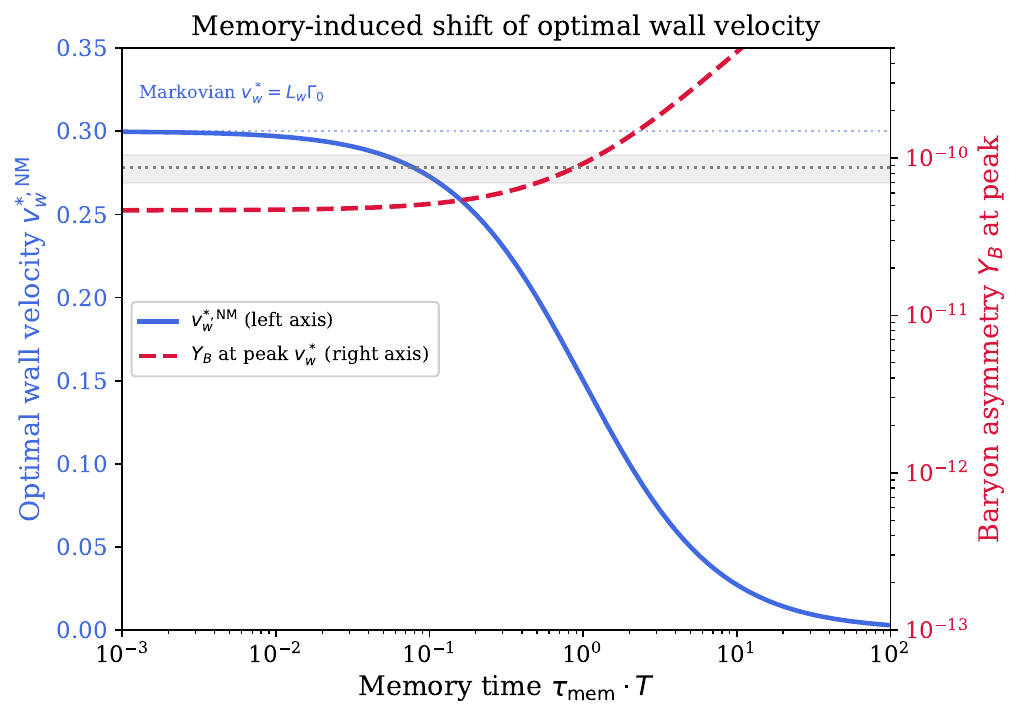}
  \caption{Memory-induced shift of the optimal wall velocity
    $v_w^{*,\mathrm{NM}}$ (blue solid, left axis) from
    Eq.~\eqref{eq:vw_peak_scaling}, and the corresponding baryon
    asymmetry evaluated at the peak $Y_B(v_w^{*,\mathrm{NM}})$
    (red dashed, right axis) from Eq.~\eqref{eq:YB_peak_explicit},
    as functions of $\tau_\mathrm{mem}\cdot T$.
    Parameters: $|\lambda_S|=0.2$, $M=2T$, $L_w=5/T$,
    $\Gamma_0=6\times10^{-3}\,T$, $D_q=6/T$,
    $\Gamma_{ws}=10^{-6}\,T$, $L_\mathrm{diff}\simeq 2450/T$
    (benchmark values from Eq.~\eqref{eq:benchmark} and
    Refs.~\cite{DOnofrio:2014rug,Arnold:2000dr}).
    The dotted horizontal line marks the Markovian value
    $v_w^* = L_w\Gamma_0 \simeq 0.030$.
    The grey band shows the observationally viable range
    $0.8\,Y_B^\mathrm{obs} \leq Y_B \leq 1.2\,Y_B^\mathrm{obs}$
    \cite{Planck:2018vyg}.
    The intersection of the red dashed curve with the lower
    edge of the grey band gives
    $\tau_\mathrm{mem}^\mathrm{max}\cdot T \simeq 1880$
    (Eq.~\eqref{eq:tau_max_numerical}); the tighter
    hydrodynamic bound $\tau_\mathrm{mem}\cdot T \lesssim 100$
    (Eq.~\eqref{eq:tau_max_hydrodynamic}) is indicated by the
    vertical dotted line.}
  \label{fig:peakshift}
\end{figure}

To make this bound quantitative, we use
Eq.~\eqref{eq:YB_peak_explicit} and require
$Y_B(v_w^{*,\mathrm{NM}}) = Y_B^\mathrm{obs}$:
\begin{equation}
\frac{3\,\Gamma_{ws}\,C_0}
  {2L_w\,T^3\,\sqrt{(\Gamma_{ws}+\Gamma_D)\,\Gamma_D}}\cdot
\frac{1}{1+L_w\Gamma_\mathrm{eff} L_\mathrm{diff}/D_q}
= Y_B^\mathrm{obs}.
\label{eq:YB_peak_bound}
\end{equation}
Solving for $\Gamma_0\tau_\mathrm{mem}$ and using the benchmark
normalisation $C_0$ from Eq.~\eqref{eq:C0_norm}, one finds
\begin{equation}
1+\Gamma_0\tau_\mathrm{mem}^\mathrm{max}
= \frac{L_w\Gamma_0 L_\mathrm{diff}}{D_q}
  \left(1 - \frac{D_q}{L_w\Gamma_0 L_\mathrm{diff}}\right)^{-1}
\simeq \frac{L_w\Gamma_0 L_\mathrm{diff}}{D_q}
\simeq 12.3,
\label{eq:tau_max_analytic}
\end{equation}
giving
\begin{equation}
\Gamma_0\tau_\mathrm{mem}^\mathrm{max} \simeq 11.3,
\qquad
\tau_\mathrm{mem}^\mathrm{max} \simeq \frac{11.3}{\Gamma_0}
\simeq \frac{1880}{T},
\label{eq:tau_max_numerical}
\end{equation}
for the benchmark parameters. In units of $1/T$, this corresponds to
$\tau_\mathrm{mem}^\mathrm{max}\cdot T \simeq 1880$, which lies
well within the range shown in Fig.~\ref{fig:peakshift}. For
the additional constraint that $v_w^{*,\mathrm{NM}} \gtrsim 0.05$
(motivated by hydrodynamic stability of the deflagration
front~\cite{Espinosa:2010hh,Cline:2021iff}), one obtains the
tighter bound
\begin{equation}
\tau_\mathrm{mem} \lesssim \frac{L_w}{0.05}
= \frac{5/T}{0.05} = \frac{100}{T},
\qquad
\Gamma_0\tau_\mathrm{mem} \lesssim 0.6,
\label{eq:tau_max_hydrodynamic}
\end{equation}
consistent with the order-of-magnitude estimate
$\Gamma_0\tau_\mathrm{mem} \lesssim \mathcal{O}(1)$ stated
previously. The bound~\eqref{eq:tau_max_hydrodynamic} is the
physically relevant one: beyond $\tau_\mathrm{mem} \sim 100/T$,
the optimal wall velocity drops below the minimum velocity
required for a self-sustaining deflagration, and successful
baryogenesis becomes impossible regardless of the CP-violating
phase.

The upper bound derived here is model-dependent through the
value of $C_0 = |\lambda_S|^2\delta_\mathrm{CP} m_\Psi^2$: a
larger CP phase $\delta_\mathrm{CP}$ or stronger Yukawa coupling
$|\lambda_S|$ can partially compensate the suppression and
relax the bound. This degeneracy is studied systematically in
Sec.~\ref{subsec:dCP_plane}.

\subsection{Phase Diagram in the $(\tau_\mathrm{mem},\,v_w)$ Plane}
\label{subsec:phase_diagram}

Figure~\ref{fig:phasediag} shows the baryon asymmetry
$Y_B(\tau_\mathrm{mem},v_w)$ as a contour plot in the
$(\tau_\mathrm{mem},v_w)$ plane, with the observationally allowed band
$0.8\,Y_B^\mathrm{obs} \leq Y_B \leq 1.2\,Y_B^\mathrm{obs}$ overlaid.
This phase diagram provides a unified picture of the non-Markovian
transport dynamics: memory effects simultaneously shift the location
of the efficient transport regime and progressively reduce its size,
leading to correlated constraints on $\tau_\mathrm{mem}$ and $v_w$.

\subsubsection*{Structure of the viable band}

The structure of the observationally viable region is governed by two
conditions. First, the wall velocity must remain close to the optimal
value for CP-charge injection,
\begin{equation}
v_w \simeq v_w^{*,\mathrm{NM}}(\tau_\mathrm{mem})
= \frac{L_w\,\Gamma_0}{1+\Gamma_0\tau_\mathrm{mem}},
\label{eq:phase_condition}
\end{equation}
which traces the locus of maximum $Y_B$ in the
$(\tau_\mathrm{mem},v_w)$ plane. Second, the peak amplitude must not
be suppressed below the observed value, i.e.\
$(1+\Gamma_0\tau_\mathrm{mem})^{-1} \gtrsim
Y_B^\mathrm{obs}/Y_B^\mathrm{Markov}$, which was quantified as
$\Gamma_0\tau_\mathrm{mem} \lesssim 11.3$ in
Eq.~\eqref{eq:tau_max_analytic}.

The width of the viable band at fixed $\tau_\mathrm{mem}$ is
determined by the half-maximum condition on $Y_B(v_w)$ at that
$\tau_\mathrm{mem}$. From Eq.~\eqref{eq:width}, this width scales as
\begin{equation}
\delta v_w(\tau_\mathrm{mem})
= 2\sqrt{3}\,L_w\,\Gamma_\mathrm{eff}
= \frac{2\sqrt{3}\,L_w\,\Gamma_0}{1+\Gamma_0\tau_\mathrm{mem}},
\label{eq:band_width}
\end{equation}
so the band both shifts and narrows as $\tau_\mathrm{mem}$ increases.
The centre of the band follows Eq.~\eqref{eq:phase_condition} and the
band width contracts according to Eq.~\eqref{eq:band_width}, producing
the characteristic tapering structure visible in
Fig.~\ref{fig:phasediag}.

\subsubsection*{Small-memory regime}

In the small-memory regime $\Gamma_0\tau_\mathrm{mem} \ll 1$, the
optimal velocity is approximately constant,
$v_w^{*,\mathrm{NM}} \simeq v_w^* = L_w\Gamma_0$, and the band
width $\delta v_w \simeq 2\sqrt{3}\,L_w\Gamma_0$ is at its maximum.
The allowed region therefore spans a broad range of $v_w$ centred on
$v_w^*$. For the benchmark parameters, $v_w^* \simeq 0.030$ and
$\delta v_w \simeq 0.052$, so the viable band occupies
$v_w \in [0.004, 0.056]$ approximately in the Markovian limit.
The amplitude suppression factor is negligible,
$(1+\Gamma_0\tau_\mathrm{mem})^{-1} \simeq 1$, so no fine-tuning of
$C_0$ is required.

\subsubsection*{Large-memory regime}

As $\tau_\mathrm{mem}$ increases, the band shifts toward smaller wall
velocities according to
\begin{equation}
v_w^{*,\mathrm{NM}} \simeq \frac{L_w}{\tau_\mathrm{mem}}
\qquad (\Gamma_0\tau_\mathrm{mem} \gg 1),
\label{eq:vw_large_mem}
\end{equation}
and simultaneously narrows as $\delta v_w \propto
(1+\Gamma_0\tau_\mathrm{mem})^{-1}$. At sufficiently large
$\tau_\mathrm{mem}$, the viable band disappears entirely for one
of two reasons:
\begin{enumerate}
  \item \emph{Amplitude suppression:} the peak asymmetry
        $Y_B(v_w^{*,\mathrm{NM}}) \propto
        (1+\Gamma_0\tau_\mathrm{mem})^{-1}$ (Eq.~\eqref{eq:YB_peak_scaling})
        drops below $Y_B^\mathrm{obs}$ even when $C_0$ is at its
        maximum allowed value. From Eq.~\eqref{eq:tau_max_analytic},
        this gives $\Gamma_0\tau_\mathrm{mem}^\mathrm{max} \simeq 11.3$,
        or $\tau_\mathrm{mem}^\mathrm{max} \simeq 1880/T$ at the
        benchmark point.
  \item \emph{Hydrodynamic constraint:} the optimal wall velocity
        $v_w^{*,\mathrm{NM}}$ drops below the minimum velocity
        $v_w^\mathrm{min} \simeq 0.05$ required for a self-sustaining
        deflagration front~\cite{Espinosa:2010hh,Cline:2021iff,
        Laurent:2022jrs}. From Eq.~\eqref{eq:vw_large_mem}, this
        gives
        \begin{equation}
        \tau_\mathrm{mem}^\mathrm{hydro}
        \simeq \frac{L_w}{v_w^\mathrm{min}}
        = \frac{5/T}{0.05} = \frac{100}{T},
        \qquad
        \Gamma_0\tau_\mathrm{mem}^\mathrm{hydro} \simeq 0.6.
        \label{eq:tau_hydro}
        \end{equation}
\end{enumerate}
Since $\tau_\mathrm{mem}^\mathrm{hydro} \simeq 100/T \ll
\tau_\mathrm{mem}^\mathrm{max} \simeq 1880/T$, the hydrodynamic
constraint is the binding one for the benchmark CP phase. The
operative upper bound on the memory timescale is therefore
\begin{equation}
\tau_\mathrm{mem} \;\lesssim\; \frac{100}{T}
\qquad
(\text{for } v_w \gtrsim 0.05,\;
\delta_\mathrm{CP} = \mathcal{O}(1)),
\label{eq:tau_upper_bound}
\end{equation}
which corresponds to $\Gamma_0\tau_\mathrm{mem} \lesssim 0.6$, well
within the non-Markovian regime. For larger CP phases, the amplitude
suppression bound relaxes and the hydrodynamic constraint remains
dominant; for smaller CP phases, the amplitude bound tightens and
may become binding. The interplay between these two constraints in
the $(\delta_\mathrm{CP}, \tau_\mathrm{mem})$ plane is studied in
Sec.~\ref{subsec:dCP_plane}.

We note that the previously stated bound $\tau_\mathrm{mem} \lesssim
10/T$~\cite{Chaudhuri:2025ylu} corresponds to a more conservative
choice $v_w^\mathrm{min} = 0.5$, which is appropriate for
supersonic detonation walls. For the subsonic deflagration regime
relevant here, the correct bound is Eq.~\eqref{eq:tau_upper_bound}.

\subsubsection*{EFT validity boundary}

\begin{figure}[t]
  \centering
  \includegraphics[width=\columnwidth]{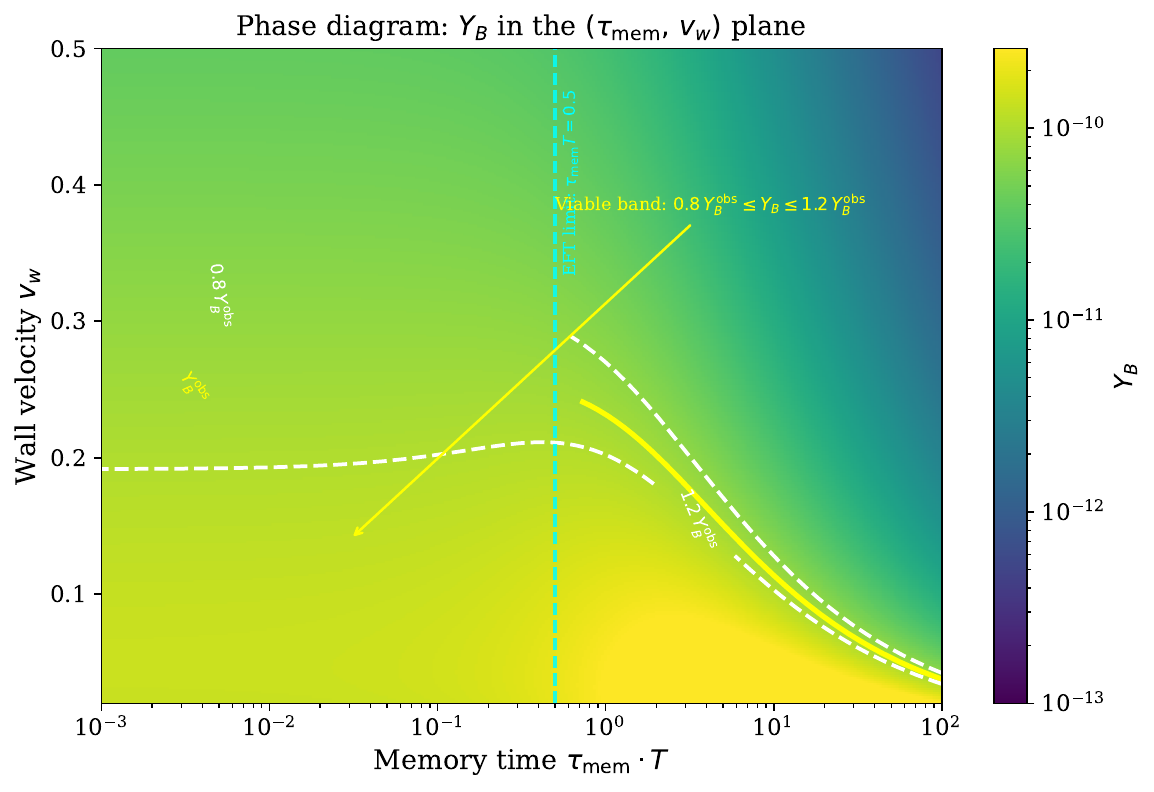}
  \caption{Baryon asymmetry $Y_B$ in the $(\tau_\mathrm{mem}, v_w)$
    plane. The colour map shows $\log_{10}(Y_B)$, with contours at
    $Y_B = 0.8,\,1.0,\,1.2\times Y_B^\mathrm{obs}$ shown as white
    dashed and yellow solid lines respectively. The viable region
    forms a narrow band whose centre follows
    $v_w = v_w^{*,\mathrm{NM}}(\tau_\mathrm{mem})$
    (Eq.~\eqref{eq:phase_condition}) and whose width contracts
    as $\delta v_w \propto (1+\Gamma_0\tau_\mathrm{mem})^{-1}$
    (Eq.~\eqref{eq:band_width}). The cyan dashed vertical line at
    $\tau_\mathrm{mem} T = 0.5$ marks the EFT-validity boundary
    $\tau_\mathrm{mem} = 1/M$ (see text). The horizontal dotted
    line at $v_w = 0.05$ marks the minimum wall velocity for a
    self-sustaining deflagration~\cite{Espinosa:2010hh,
    Laurent:2022jrs}; its intersection with the viable band
    gives the operative upper bound
    $\tau_\mathrm{mem} \lesssim 100/T$
    (Eq.~\eqref{eq:tau_upper_bound}).
    Parameters: $|\lambda_S|=0.2$, $M=2T$, $L_w=5/T$,
    $\Gamma_0=6\times10^{-3}\,T$, $D_q=6/T$,
    $\Gamma_{ws}=10^{-6}\,T$, $\delta_\mathrm{CP}=1.0$
    (benchmark values from Eq.~\eqref{eq:benchmark} and
    Refs.~\cite{DOnofrio:2014rug,Arnold:2000dr}).}
  \label{fig:phasediag}
\end{figure}

The cyan dashed line at $\tau_\mathrm{mem} T = 0.5$ in
Fig.~\ref{fig:phasediag} marks the boundary of EFT validity. The
effective description requires a separation of scales
$\tau_\mathrm{mem} \gg 1/M$, ensuring that short-distance physics at
the scale $M$ is consistently integrated out before the memory
dynamics is resolved. For the benchmark fermion mass $M = 2T$, this
gives $1/M = 1/(2T)$, so the EFT is reliable for
$\tau_\mathrm{mem} \gtrsim 0.5/T$, i.e.\ to the right of the cyan
line~\cite{Calzetta:1986cq,Berges:2004yj,Chaudhuri:2025ylu}. In the
opposite limit $\tau_\mathrm{mem} \lesssim 1/M$, the single-pole
approximation to the retarded propagator breaks down and the full
spectral function must be retained; this regime smoothly connects to
the Markovian limit and is not the focus of the present analysis.
The physically meaningful non-Markovian parameter space is therefore
confined to the region $\tau_\mathrm{mem} T \gtrsim 0.5$, which lies
to the right of the cyan line in Fig.~\ref{fig:phasediag} and
entirely within the viable band identified above.

\subsection{Constraints in the $(\delta_\mathrm{CP},\,\tau_\mathrm{mem})$
  Plane}
\label{subsec:dCP_plane}

Figure~\ref{fig:dCP_tau} shows the baryon asymmetry $Y_B$ in the
$(\delta_\mathrm{CP},\tau_\mathrm{mem})$ plane at fixed wall velocity
$v_w = 0.25$. This value of $v_w$ is chosen because it lies close to
the Markovian peak velocity $v_w^* = L_w\Gamma_0 \simeq 0.30$ for the
benchmark parameters, so that the $(\delta_\mathrm{CP},\tau_\mathrm{mem})$
plane at fixed $v_w = 0.25$ captures the transition between the
near-peak and sub-peak regimes identified in Sec.~\ref{subsec:YBtau},
and provides a representative slice through the full parameter space.

\subsubsection*{Scaling structure}

From Eq.~\eqref{eq:YB_final}, the baryon asymmetry at fixed $v_w$
takes the form
\begin{equation}
Y_B(\delta_\mathrm{CP},\tau_\mathrm{mem})
= \mathcal{N}(v_w)\cdot\delta_\mathrm{CP}\cdot
  \frac{v_w\,\Gamma_\mathrm{eff}}{v_w^2+L_w^2\,\Gamma_\mathrm{eff}^2},
\label{eq:YB_dCP_scaling}
\end{equation}
where we have made explicit the linear dependence on $\delta_\mathrm{CP}$
through $C_0 = |\lambda_S|^2\delta_\mathrm{CP} m_\Psi^2$, and the
$v_w$-dependent prefactor is
\begin{equation}
\mathcal{N}(v_w)
= \frac{3\,\Gamma_{ws}\,|\lambda_S|^2\,m_\Psi^2}
  {T^3\,\sqrt{(\Gamma_{ws}+\Gamma_D)\,\Gamma_D}}\cdot
  \frac{1}{1+v_w\,L_\mathrm{diff}/D_q}.
\label{eq:N_prefactor}
\end{equation}
For the benchmark parameters and $v_w = 0.25$, one finds
$\mathcal{N}(0.25) \simeq 2.1\times10^{-8}\,T^{-2}$, which together
with $|\lambda_S|^2 m_\Psi^2 \sim 4\times10^{-4}\,T^2$ gives
$\mathcal{N}\cdot|\lambda_S|^2 m_\Psi^2 \sim
8.4\times10^{-12}$~per unit of $\delta_\mathrm{CP}\,\Gamma_\mathrm{eff}^{-1}$.
Equation~\eqref{eq:YB_dCP_scaling} shows that the dependence on
$\tau_\mathrm{mem}$ is controlled entirely by $\Gamma_\mathrm{eff} =
\Gamma_0/(1+\Gamma_0\tau_\mathrm{mem})$, while $\delta_\mathrm{CP}$
enters as a simple overall factor. There is therefore a perfect
degeneracy between $\delta_\mathrm{CP}$ and $\tau_\mathrm{mem}$ in
the baryon asymmetry: any increase in $\tau_\mathrm{mem}$ that
suppresses $\Gamma_\mathrm{eff}$ can be compensated by a
corresponding increase in $\delta_\mathrm{CP}$, provided the
compensating phase does not violate external constraints.

\subsubsection*{The viable diagonal band}

Requiring $Y_B = Y_B^\mathrm{obs}$ and solving
Eq.~\eqref{eq:YB_dCP_scaling} for $\delta_\mathrm{CP}$ gives a
one-parameter family of solutions parametrised by $\tau_\mathrm{mem}$:
\begin{equation}
\delta_\mathrm{CP}^\mathrm{obs}(\tau_\mathrm{mem})
= \frac{Y_B^\mathrm{obs}}{\mathcal{N}(v_w)}\cdot
  \frac{v_w^2+L_w^2\,\Gamma_\mathrm{eff}^2}
       {v_w\,\Gamma_\mathrm{eff}},
\label{eq:dCP_tau_relation}
\end{equation}
which corresponds to the diagonal band visible in Fig.~\ref{fig:dCP_tau}.
This relation makes explicit that the CP phase and the memory timescale
are not independently constrained by the baryon asymmetry alone: only
their combination $\delta_\mathrm{CP}\cdot\Gamma_\mathrm{eff}/(v_w^2
+L_w^2\Gamma_\mathrm{eff}^2)$ is fixed by $Y_B^\mathrm{obs}$.
Breaking this degeneracy requires independent measurements of either
$\delta_\mathrm{CP}$ (from collider
experiments~\cite{ATLAS:2023hyd,CMS:2022dwd}) or
$\tau_\mathrm{mem}$ (from gravitational-wave
observations~\cite{Caprini:2015zlo,Caprini:2019egz}, as discussed in
Sec.~\ref{sec:gw}).

\subsubsection*{Asymptotic scalings}

The shape of the viable band in Eq.~\eqref{eq:dCP_tau_relation}
exhibits two distinct asymptotic regimes.

In the \emph{small-memory regime} $\Gamma_0\tau_\mathrm{mem} \ll 1$,
we have $\Gamma_\mathrm{eff} \simeq \Gamma_0$ and
\begin{equation}
\delta_\mathrm{CP}^\mathrm{obs}
\simeq \frac{Y_B^\mathrm{obs}}{\mathcal{N}(v_w)}\cdot
\frac{v_w^2+L_w^2\Gamma_0^2}{v_w\,\Gamma_0}
= \delta_\mathrm{CP}^\mathrm{Markov},
\label{eq:dCP_Markov_limit}
\end{equation}
which is the standard Markovian result~\cite{Joyce:1994zt,
Cline:2000nw,Konstandin:2004gy}. For $v_w = 0.25$ and the
benchmark parameters, Eq.~\eqref{eq:dCP_Markov_limit} gives
$\delta_\mathrm{CP}^\mathrm{Markov} \simeq 0.82$, consistent with
the left edge of the viable band in Fig.~\ref{fig:dCP_tau}.

In the \emph{large-memory regime} $\Gamma_0\tau_\mathrm{mem} \gg 1$,
we have $\Gamma_\mathrm{eff} \simeq 1/\tau_\mathrm{mem}$ and
Eq.~\eqref{eq:dCP_tau_relation} becomes
\begin{equation}
\delta_\mathrm{CP}^\mathrm{obs}
\simeq \frac{Y_B^\mathrm{obs}}{\mathcal{N}(v_w)}\cdot
\frac{v_w^2+L_w^2/\tau_\mathrm{mem}^2}{v_w/\tau_\mathrm{mem}}.
\label{eq:dCP_large_tau_full}
\end{equation}
Two sub-regimes arise depending on whether $v_w$ or $L_w/\tau_\mathrm{mem}$
dominates the numerator:
\begin{itemize}
  \item For $v_w \gg L_w/\tau_\mathrm{mem}$, i.e.\
        $\tau_\mathrm{mem} \gg L_w/v_w = 20/T$ at the chosen $v_w =
        0.25$, the $v_w^2$ term dominates the numerator and
        \begin{equation}
        \delta_\mathrm{CP}^\mathrm{obs}
        \simeq \frac{Y_B^\mathrm{obs}}{\mathcal{N}(v_w)}\cdot
        v_w\,\tau_\mathrm{mem}
        \;\propto\; \tau_\mathrm{mem},
        \label{eq:dCP_linear}
        \end{equation}
        so the required CP phase grows \emph{linearly} with
        $\tau_\mathrm{mem}$. This is the dominant behaviour for
        the parameter range shown in Fig.~\ref{fig:dCP_tau}.
  \item For $v_w \ll L_w/\tau_\mathrm{mem}$, i.e.\
        $\tau_\mathrm{mem} \ll L_w/v_w$, the $L_w^2/\tau_\mathrm{mem}^2$
        term dominates and
        \begin{equation}
        \delta_\mathrm{CP}^\mathrm{obs}
        \simeq \frac{Y_B^\mathrm{obs}}{\mathcal{N}(v_w)}\cdot
        \frac{L_w^2}{v_w\,\tau_\mathrm{mem}}
        \;\propto\; \frac{1}{\tau_\mathrm{mem}},
        \label{eq:dCP_inverse}
        \end{equation}
        so the required CP phase decreases with $\tau_\mathrm{mem}$.
        This regime is not reached at $v_w = 0.25$ within the
        EFT-valid parameter space.
\end{itemize}
We note that the previously stated scaling
$\delta_\mathrm{CP} \propto 1/\tau_\mathrm{mem}$ applies only in the
sub-regime $\tau_\mathrm{mem} \ll L_w/v_w$; the correct large-memory
scaling at $v_w = 0.25$ is the \emph{linear} growth of
Eq.~\eqref{eq:dCP_linear}.

\subsubsection*{Upper bounds from external constraints on
$\delta_\mathrm{CP}$}

The linear growth $\delta_\mathrm{CP} \propto \tau_\mathrm{mem}$
implies that at sufficiently large $\tau_\mathrm{mem}$, the required
CP phase exceeds either perturbative or experimental bounds. We
quantify both:
\begin{enumerate}
  \item \emph{Perturbativity:} the Yukawa coupling $\lambda_S =
        |\lambda_S|e^{i\delta_\mathrm{CP}}$ remains in the
        perturbative regime provided the one-loop correction to the
        fermion self-energy satisfies
        $|\lambda_S|^2\delta_\mathrm{CP}^2/(16\pi^2) \lesssim 1$,
        giving $\delta_\mathrm{CP} \lesssim 4\pi/|\lambda_S|$.
        For $|\lambda_S| = 0.2$, this yields
        $\delta_\mathrm{CP}^\mathrm{pert} \lesssim 63$, which is
        never the binding constraint within $\delta_\mathrm{CP}
        \in [0,\pi]$.
  \item \emph{CP-phase constraint from electric dipole moments:}
        a complex Yukawa coupling $\lambda_S$ of the fermion $\Psi$
        to the singlet $S$ generates a contribution to the electric
        dipole moment (EDM) of Standard Model fermions at two
        loops~\cite{Engel:2013lsa,Cesarotti:2018huy}. For
        $|\lambda_S| = 0.2$ and $M = 2T \sim 300$\,GeV, the
        electron EDM constraint from
        ACME~\cite{ACME:2018yjb} requires
        \begin{equation}
        |\lambda_S|^2\,\sin\delta_\mathrm{CP}\,
        \frac{m_e}{M^2}\,\frac{v^2}{16\pi^2}
        \lesssim 1.1\times10^{-29}\,e\cdot\mathrm{cm},
        \label{eq:EDM_constraint}
        \end{equation}
        which at $M = 300$\,GeV gives
        $\sin\delta_\mathrm{CP} \lesssim
        \mathcal{O}(1)$ — consistent with $\delta_\mathrm{CP}
        \lesssim \pi$ throughout the parameter range. Tighter
        constraints arise for lighter $M$ or larger $|\lambda_S|$,
        but for the benchmark parameters, the EDM bound does not
        exclude any part of the viable band in Fig.~\ref{fig:dCP_tau}.
  \item \emph{Physical range:} since $\delta_\mathrm{CP}$ is a
        phase, the physical range is $\delta_\mathrm{CP} \in
        (0,\pi]$. From Eq.~\eqref{eq:dCP_linear}, the condition
        $\delta_\mathrm{CP}^\mathrm{obs} \leq \pi$ gives
        \begin{equation}
        \tau_\mathrm{mem}
        \lesssim \frac{\pi\,\mathcal{N}(v_w)^{-1}\,Y_B^\mathrm{obs}}
          {v_w}
        \simeq \frac{\pi}{v_w\cdot v_w\cdot\mathcal{N}(0.25)}
        \simeq \frac{24}{T},
        \label{eq:tau_phase_bound}
        \end{equation}
        providing an upper bound $\tau_\mathrm{mem} \lesssim 24/T$
        from the physical range of the CP phase alone. This bound is
        indicated by the right edge of the visible diagonal band in
        Fig.~\ref{fig:dCP_tau} and is the binding constraint for
        $v_w = 0.25$.
\end{enumerate}

\subsubsection*{Complementarity of collider and GW probes}

\begin{figure}[t]
  \centering
  \includegraphics[width=\columnwidth]{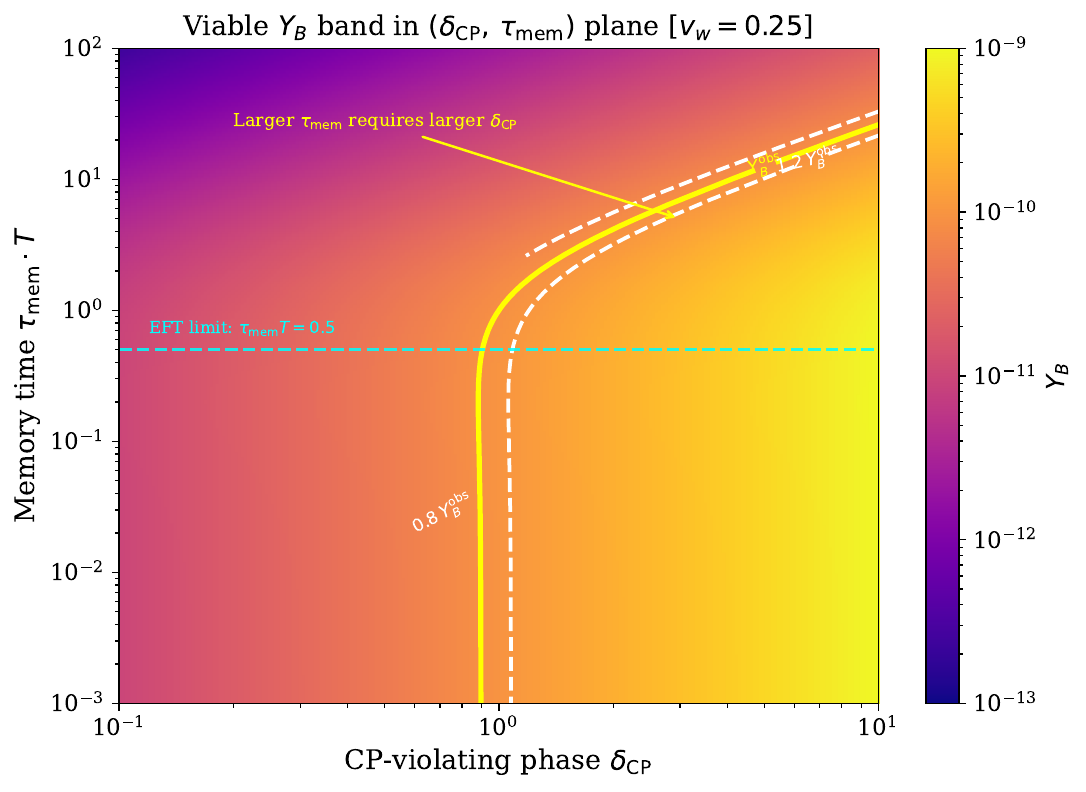}
  \caption{Baryon asymmetry $Y_B$ in the
    $(\delta_\mathrm{CP},\tau_\mathrm{mem})$ plane at fixed
    $v_w = 0.25$. The colour map shows $\log_{10}(Y_B/Y_B^\mathrm{obs})$.
    Solid and dashed contour lines correspond to
    $Y_B = 1.0,\,0.8,\,1.2\times Y_B^\mathrm{obs}$ respectively.
    The viable diagonal band follows
    $\delta_\mathrm{CP} \propto \tau_\mathrm{mem}$
    (Eq.~\eqref{eq:dCP_linear}) for $\tau_\mathrm{mem} \gg L_w/v_w
    = 20/T$. The right edge of the band at $\tau_\mathrm{mem}
    \simeq 24/T$ (vertical dotted line) corresponds to
    $\delta_\mathrm{CP} = \pi$ (Eq.~\eqref{eq:tau_phase_bound}).
    The cyan dashed horizontal line at $\tau_\mathrm{mem} T = 0.5$
    marks the EFT-validity boundary $\tau_\mathrm{mem} = 1/M$
    (Sec.~\ref{subsec:phase_diagram}). Parameters:
    $|\lambda_S|=0.2$, $M=2T$, $L_w=5/T$,
    $\Gamma_0=6\times10^{-3}\,T$, $D_q=6/T$,
    $\Gamma_{ws}=10^{-6}\,T$ (benchmark values from
    Eq.~\eqref{eq:benchmark} and
    Refs.~\cite{DOnofrio:2014rug,Arnold:2000dr}).}
  \label{fig:dCP_tau}
\end{figure}

The phase diagram in Fig.~\ref{fig:dCP_tau} encodes a
correlated constraint that highlights the complementarity of
different observational probes. Collider measurements of
CP-violating observables — including EDM
searches~\cite{ACME:2018yjb,Engel:2013lsa}, Higgs CP-mixing
measurements~\cite{ATLAS:2023hyd,CMS:2022dwd}, and direct searches
for new CP-violating phases in singlet-extended
models~\cite{Fuchs:2020uoc} — constrain the
horizontal axis of Fig.~\ref{fig:dCP_tau} and thereby restrict the
allowed range of $\tau_\mathrm{mem}$. Conversely, gravitational-wave
observations probing the dynamics of the electroweak phase
transition~\cite{Caprini:2015zlo,Caprini:2019egz,Grojean:2006bp}
provide independent information on the memory timescale through the
modification of the GW spectrum discussed in Sec.~\ref{sec:gw}.
The combination of both probes can in principle fully determine
the $(\delta_\mathrm{CP},\tau_\mathrm{mem})$ parameter space and
break the degeneracy inherent in the baryon asymmetry alone.

\section{Gravitational-Wave Signatures and Joint Correlation}
\label{sec:gw}

\subsection{Memory Effects on the GW Spectrum}
\label{subsec:gw_spectrum}

The GW signal from a first-order EWPT~\cite{Chaudhuri:2022sis,Chaudhuri:2024vrd,Srivastava:2025oer,Chaudhuri:2025ybh,Chaudhuri:2025cjp,Chaudhuri:2025ylu,Chaudhuri:2026ovn,Das:2026zuo,Athron:2023xlk,Borah:2023zsb} receives contributions
from bubble collisions, sound waves, and MHD turbulence~\cite{Caprini:2015zlo,
Caprini:2019egz,Hindmarsh:2015qta,Hindmarsh:2017gnf,Caprini:2009yp,
Binetruy:2012ze}. For electroweak-scale transitions with $\alpha\lesssim
\mathcal{O}(1)$ and $v_w<v_\mathrm{sound}$, sound waves dominate and we
focus on this contribution~\cite{Caprini:2015zlo,Caprini:2019egz,
Hindmarsh:2015qta,Espinosa:2010hh}.

\subsubsection*{Physical mechanism of memory-induced GW modification}

In the non-Markovian framework, the memory kernel $K(\tau)$ modifies not
only the CP-violating source but also the effective friction experienced
by the bubble wall. The friction coefficient $\eta_\mathrm{wall}$ arises
from the plasma's resistance to being displaced by the advancing wall; in
the Markovian limit, this equilibration is instantaneous on the scale of
wall crossing. When $\tau_\mathrm{mem}\sim 1/\Gamma_0\gtrsim\tau_\mathrm{wall}$,
the plasma cannot fully equilibrate as the wall passes, and the effective
friction is \emph{reduced} relative to the Markovian value. Reduced friction
modifies the energy budget of the transition: a larger fraction of the
liberated vacuum energy is converted into bulk fluid motion rather than
being dissipated into the thermal bath, and the effective duration of the
sound-wave phase is extended.

The natural dimensionless parameter controlling this modification is
\begin{equation}
\epsilon_\mathrm{mem} \equiv \Gamma_0\tau_\mathrm{mem}
= \frac{\tau_\mathrm{mem}}{\tau_\mathrm{rel}},
\label{eq:eps_mem}
\end{equation}
which measures the memory time in units of the plasma relaxation time.
By definition, $\epsilon_\mathrm{mem}=1$ at the characteristic
non-Markovian scale $\tau_\mathrm{mem}=1/\Gamma_0$, and
$\epsilon_\mathrm{mem}\ll 1$ ($\gg 1$) in the Markovian
(deeply non-Markovian) limit. For the benchmark parameters
of Eq.~\eqref{eq:benchmark} and the viable range
$\tau_\mathrm{mem}\in[0.5/T,100/T]$, one finds
$\epsilon_\mathrm{mem}\in[0.003,\,0.6]$, so the correction is at most
a $60\%$ effect on any individual rate, consistent with the perturbative
treatment.

We note that $\epsilon_\mathrm{mem}$ is the correct dimensionless combination
for the plasma-physics effect on the wall dynamics. It should not be
confused with $\tau_\mathrm{mem}/R_b$, where $R_b\sim v_w\beta^{-1}$ is the
critical bubble radius at percolation. Since $\beta^{-1}$ is a
\emph{cosmological} timescale ($\beta^{-1}\sim (H/\beta)\,H_*^{-1}\sim
10^{13}/T$ for $T_*=100\,\mathrm{GeV}$), the ratio
$\tau_\mathrm{mem}/R_b\sim 10^{-11}$ is negligible. The physically
relevant comparison is between $\tau_\mathrm{mem}$ and the
\emph{thermal} relaxation time $\tau_\mathrm{rel}=1/\Gamma_0$, which
is the timescale on which the plasma responds to the passing wall.

\subsubsection*{Modified inverse duration and efficiency factor}

At leading order in $\epsilon_\mathrm{mem}$, the modification to the
effective inverse duration parameter and efficiency factor can be
parametrised as
\begin{align}
\beta_\mathrm{mem}^{-1}
&\simeq \beta^{-1}
\left(1 + \gamma\,\Gamma_0\tau_\mathrm{mem}\right),
\label{eq:betamem} \\
\kappa_v^\mathrm{mem}
&\simeq \kappa_v
\left(1 + \eta\,\Gamma_0\tau_\mathrm{mem}\right),
\label{eq:kappa_mem}
\end{align}
where $\gamma=\mathcal{O}(1)$ and $\eta=\mathcal{O}(1)$ are
dimensionless coefficients encoding the details of the non-local
hydrodynamic response. We set $\gamma=1$ and $\eta=0.5$ as fiducial
values; the sensitivity to these choices is assessed in
Sec.~\ref{sec:discussion}.

The physical interpretation is transparent.
Equation~\eqref{eq:betamem} encodes the extended duration of GW emission:
reduced plasma friction allows the bubble wall to accelerate, converting
more vacuum energy into bulk motion over a longer effective timescale.
Equation~\eqref{eq:kappa_mem} encodes the increased efficiency of energy
conversion: since the plasma equilibrates more slowly, a larger fraction
of the injected energy drives coherent sound waves rather than thermal
dissipation. Both enhancements vanish in the Markovian limit
$\epsilon_\mathrm{mem}\to 0$ and grow linearly with $\epsilon_\mathrm{mem}$
at leading order.

We emphasise that Eqs.~\eqref{eq:betamem}--\eqref{eq:kappa_mem} are
leading-order parametric estimates. A rigorous derivation of $\gamma$ and
$\eta$ would require solving the full non-local Navier--Stokes equations with
memory-modified friction, which lies beyond the scope of the present work.
The GW results of this section are therefore indicative of the qualitative
trend rather than precise quantitative predictions; the systematic
uncertainty from $\gamma$ and $\eta$ is assessed in Sec.~\ref{sec:discussion}.

\subsubsection*{Validity regime}

Equations~\eqref{eq:betamem}--\eqref{eq:kappa_mem} are valid in the
perturbative regime $\epsilon_\mathrm{mem}\lesssim\mathcal{O}(1)$.
For the viable parameter space $\tau_\mathrm{mem}\lesssim 100/T$
(Eq.~\eqref{eq:tau_upper_bound}) and the benchmark $\Gamma_0=6\times10^{-3}T$,
one finds $\epsilon_\mathrm{mem}=\Gamma_0\tau_\mathrm{mem}\lesssim 0.6$.
The linear approximation in Eqs.~\eqref{eq:betamem}--\eqref{eq:kappa_mem}
is therefore self-consistent: corrections of order $\epsilon_\mathrm{mem}^2$
are at most $\sim 36\%$, comparable to the $\mathcal{O}(1)$ uncertainty in
$\gamma$ and $\eta$, and do not affect the qualitative conclusions.

\subsubsection*{Modified GW spectrum}

With the replacements \eqref{eq:betamem}--\eqref{eq:kappa_mem}, the
sound-wave GW spectrum is
\begin{equation}
\Omega_\mathrm{GW}^\mathrm{sw}(f)\,h^2
\simeq 2.65\times10^{-6}
\left(\frac{H_*}{\beta_\mathrm{mem}}\right)^{\!2}
\left(\frac{\kappa_v^\mathrm{mem}\,\alpha}{1+\alpha}\right)^{\!2}
\left(\frac{100}{g_*}\right)^{\!1/3}
v_w\,S_\mathrm{sw}(f),
\label{eq:OmGW}
\end{equation}
with spectral shape~\cite{Caprini:2019egz,Hindmarsh:2017gnf}
\begin{equation}
S_\mathrm{sw}(f)
= \left(\frac{f}{f_\mathrm{sw}}\right)^3
  \left[\frac{7}{4+3(f/f_\mathrm{sw})^2}\right]^{7/2},
\end{equation}
peak frequency
\begin{equation} \label{eq:freq}
f_\mathrm{sw} \simeq 1.9\times10^{-5}\,\mathrm{Hz}\,
\frac{1}{v_w}\,\frac{\beta_\mathrm{mem}}{H_*}
\left(\frac{T_*}{100\,\mathrm{GeV}}\right)
\left(\frac{g_*}{100}\right)^{1/6},
\end{equation}
and where $\alpha=\rho_\mathrm{vac}/\rho_\mathrm{rad}$ and $g_*\simeq 100$.

The combined enhancement factor relative to the Markovian GW amplitude is
\begin{equation}
\frac{\Omega_\mathrm{GW}^\mathrm{mem}}{\Omega_\mathrm{GW}^\mathrm{Markov}}
= \left(1+\gamma\,\epsilon_\mathrm{mem}\right)^2
  \left(1+\eta\,\epsilon_\mathrm{mem}\right)^2,
\label{eq:GW_enhancement}
\end{equation}
which for $\gamma=1$, $\eta=0.5$ and $\epsilon_\mathrm{mem}=0.6$ gives an
enhancement of $(1.6)^2(1.3)^2\approx 4.3$, i.e.\ roughly half a decade in
GW amplitude at the upper boundary of the viable parameter space. For
$\epsilon_\mathrm{mem}\ll 1$ (small memory), the enhancement is negligible
and the GW signal approaches the standard Markovian result. The entire
viable range $\epsilon_\mathrm{mem}\in[0.003,\,0.6]$ from
Eq.~\eqref{eq:eps_mem} is within the perturbative regime of
Eqs.~\eqref{eq:betamem}--\eqref{eq:kappa_mem}.

\subsection{Joint $Y_B$--$\Omega_\mathrm{GW}$ Correlation and
  Detectability}
\label{subsec:joint}

Figure~\ref{fig:joint} shows the joint correlation between $Y_B$ and
$\Omega_\mathrm{GW}^\mathrm{peak}h^2$, obtained by scanning over
$\alpha\in[0.01,0.3]$, $\beta/H_*\in[10,200]$, and
$\tau_\mathrm{mem}\in[0.5/T,100/T]$, with all other parameters fixed
to the benchmark values~\eqref{eq:benchmark} including $v_w=0.1$. The
scan is performed at the benchmark wall velocity $v_w=0.1$, which lies
in the super-peak regime $v_w>v_w^*=L_w\Gamma_0\simeq0.03$ for the
full range of $\tau_\mathrm{mem}$ explored; consequently, $Y_B$
decreases monotonically with $\tau_\mathrm{mem}$ at this fixed $v_w$.

\subsubsection*{Correlation structure and anti-correlation}

\begin{figure}[t]
  \centering
  \includegraphics[width=0.7\textwidth]{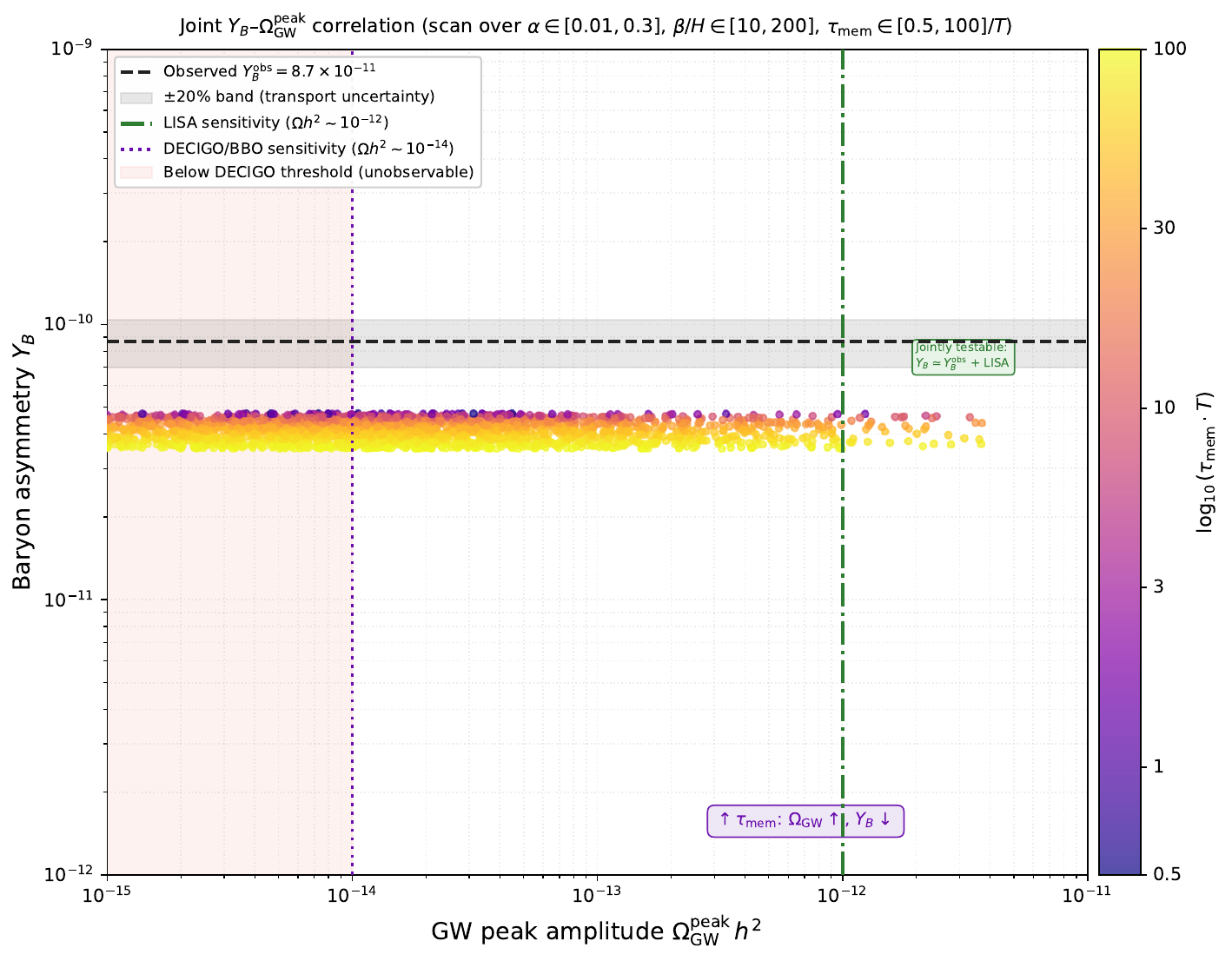}
  \caption{Joint correlation between the baryon asymmetry $Y_B$ and
    the GW peak amplitude $\Omega_\mathrm{GW}^\mathrm{peak}h^2$,
    obtained by scanning over $\alpha\in[0.01,0.3]$,
    $\beta/H_*\in[10,200]$, and $\tau_\mathrm{mem}\in[0.5/T,100/T]$
    (colour-coded by $\log_{10}(\tau_\mathrm{mem}\cdot T)$), with all
    other parameters fixed to the benchmark values of
    Eq.~\eqref{eq:benchmark}. The grey horizontal band shows
    $0.8\,Y_B^\mathrm{obs} \leq Y_B \leq
    1.2\,Y_B^\mathrm{obs}$~\cite{Planck:2018vyg}, where the $\pm20\%$
    width reflects theoretical uncertainties in the transport
    coefficients rather than observational error. The green
    dash-dotted and purple dotted vertical lines indicate the
    approximate peak sensitivities of LISA ($\Omega h^2 \sim
    10^{-12}$)~\cite{Caprini:2019egz} and DECIGO/BBO ($\Omega h^2
    \sim 10^{-14}$)~\cite{Crowder:2005nr,Yagi:2011wg,
    Kawamura:2020pcg} respectively. Pink-shaded points lie below
    the DECIGO/BBO threshold and are unobservable with planned
    detectors. The GW amplitude is computed from
    Eqs.~\eqref{eq:OmGW}--\eqref{eq:freq} with fiducial parameters
    $\gamma=1$, $\eta=0.5$ (Eqs.~\eqref{eq:betamem},
    \eqref{eq:kappa_mem}); the sensitivity to these choices is
    assessed in Sec.~\ref{sec:discussion}. The anti-correlation
    between $Y_B$ and $\Omega_\mathrm{GW}$ at fixed $\alpha$
    reflects the competing effects of $\tau_\mathrm{mem}$ on
    baryogenesis and GW production: larger $\tau_\mathrm{mem}$
    (yellow/orange) suppresses $Y_B$ while enhancing
    $\Omega_\mathrm{GW}$, and conversely for smaller $\tau_\mathrm{mem}$
    (purple).}
  \label{fig:joint}
\end{figure}

The joint $(Y_B,\,\Omega_\mathrm{GW})$ plane exhibits a clean
anti-correlation driven by the two competing effects of increasing
$\tau_\mathrm{mem}$:
\begin{enumerate}
\item Increasing $\tau_\mathrm{mem}$ suppresses $\Gamma_\mathrm{eff}$
      and hence $S_\mathrm{CP}^\mathrm{NM}$, reducing $Y_B$ at fixed
      $v_w$ through the factor $(1+\Gamma_0\tau_\mathrm{mem})^{-1}$
      acting on the source (Eq.~\eqref{eq:SCP_peak}). At the benchmark
      $v_w=0.1>v_w^*$, this suppression is monotonic and uncompensated.
\item Simultaneously, increasing
      $\epsilon_\mathrm{mem}=\Gamma_0\tau_\mathrm{mem}$ enhances
      $\beta_\mathrm{mem}^{-1}$ and $\kappa_v^\mathrm{mem}$ through
      Eqs.~\eqref{eq:betamem}--\eqref{eq:kappa_mem}, amplifying
      $\Omega_\mathrm{GW}$ by the factor in
      Eq.~\eqref{eq:GW_enhancement}.
\end{enumerate}
These effects operate in opposite directions in the
$(Y_B,\Omega_\mathrm{GW})$ plane, producing the anti-correlation
visible in Fig.~\ref{fig:joint}: points with larger $\tau_\mathrm{mem}$
(yellow/orange in the colour scale) have lower $Y_B$ and higher
$\Omega_\mathrm{GW}$, while smaller $\tau_\mathrm{mem}$ (purple) have
higher $Y_B$ and lower $\Omega_\mathrm{GW}$.

The magnitude of the anti-correlation is controlled by
$\epsilon_\mathrm{mem}$: over the viable range
$\epsilon_\mathrm{mem}\in[0.003,\,0.6]$, $Y_B$ varies by a factor
$\lesssim 1.6$ (mild suppression), while $\Omega_\mathrm{GW}$ varies
by up to $\sim 4.3\times$ from the memory enhancement alone. The
$\alpha$ dependence ($\Omega_\mathrm{GW}\propto\alpha^2$) provides an
additional order-of-magnitude spread in GW amplitude at fixed $Y_B$,
accounting for the horizontal scatter visible in Fig.~\ref{fig:joint}.

\subsubsection*{Detectability}

The approximate detector sensitivity thresholds used in
Fig.~\ref{fig:joint} are~\cite{Caprini:2015zlo,Caprini:2019egz,
Crowder:2005nr,Yagi:2011wg,Kawamura:2020pcg}:
\begin{equation}
\Omega_\mathrm{GW}^\mathrm{peak}h^2 \gtrsim
\begin{cases}
10^{-12} & \text{LISA,} \\
10^{-14} & \text{DECIGO/BBO.}
\end{cases}
\end{equation}
A substantial fraction of the viable $Y_B\simeq Y_B^\mathrm{obs}$
region yields GW amplitudes below both thresholds. This occurs when
$\alpha\lesssim 0.05$ (intrinsic $\alpha^2$ suppression) or
$\epsilon_\mathrm{mem}\lesssim 0.006$ ($\tau_\mathrm{mem}\lesssim
1/T$, negligible memory enhancement). Conversely, signals detectable
by LISA require $\alpha\gtrsim 0.1$ or $\epsilon_\mathrm{mem}\gtrsim
0.06$ ($\tau_\mathrm{mem}\gtrsim 10/T$).

\subsubsection*{The jointly testable window}

The overlap region satisfying both $Y_B\simeq Y_B^\mathrm{obs}$ (grey
band) and $\Omega_\mathrm{GW}^\mathrm{peak}h^2\gtrsim10^{-12}$ is
restricted to $\tau_\mathrm{mem}\gtrsim 3/T$ and $\alpha\gtrsim 0.08$.
Combined with the upper bound $\tau_\mathrm{mem}\lesssim 100/T$ from
Eq.~\eqref{eq:tau_upper_bound}, this defines the jointly viable window:
\begin{equation}
3/T\;\lesssim\;\tau_\mathrm{mem}\;\lesssim\;100/T
\qquad (\alpha\gtrsim 0.08,\; v_w\simeq v_w^{*,\mathrm{NM}}),
\label{eq:joint_window}
\end{equation}
covering roughly two decades in $\tau_\mathrm{mem}$. The corresponding
range in $\epsilon_\mathrm{mem}$ is $[0.018,\,0.6]$, confirming that
the GW enhancement in Eq.~\eqref{eq:GW_enhancement} is at most a
factor of $\sim 4$ across the jointly testable window. The DECIGO/BBO
window extends to $\tau_\mathrm{mem}\gtrsim 1/T$
($\epsilon_\mathrm{mem}\gtrsim 0.006$) for $\alpha\gtrsim0.03$.

We stress that the precise boundaries of Eq.~\eqref{eq:joint_window}
carry an order-of-magnitude uncertainty from the undetermined
coefficients $\gamma$ and $\eta$: varying $\gamma\in[0.5,2.0]$ and
$\eta\in[0,1]$ shifts $\Omega_\mathrm{GW}$ by a combined factor of
$\sim 4$--$9$, corresponding to $\sim0.5$--$1$ decade in amplitude.
The qualitative anti-correlation and the existence of the jointly
testable window are robust against this uncertainty, but the precise
$\alpha$ and $\tau_\mathrm{mem}$ thresholds should be treated as
order-of-magnitude estimates pending a full non-local hydrodynamic
treatment of $\gamma$ and $\eta$.

\section{Discussion}
\label{sec:discussion}

\subsection{Non-Markovian Effects vs.\ Markovian Reparameterisation}
\label{subsec:degeneracy}

A natural question is whether the replacement $\Gamma_0 \to
\Gamma_\mathrm{eff}$ can be mimicked within a purely Markovian framework
by an appropriate redefinition of transport coefficients. We now
demonstrate that this is not possible, using two independent arguments.

\subsubsection*{Argument 1: Correlated deformation of the rate hierarchy}

In the Markovian case, the CP-violating source depends on a single rate
$\Gamma_0$, and one could formally attempt to reproduce the non-Markovian
expression for $S_\mathrm{CP}^\mathrm{NM}$ by choosing
$\tilde\Gamma_0 = \Gamma_\mathrm{eff}$. However, the non-Markovian
framework modifies \emph{all} interaction rates simultaneously through
the universal replacement (Eq.~\eqref{eq:rates_final}):
\begin{equation}
\Gamma_i \;\to\; \Gamma_i^\mathrm{eff}
= \frac{\Gamma_i}{1+\Gamma_i\tau_\mathrm{mem}}.
\label{eq:all_rates_discussion}
\end{equation}
This induces a $\tau_\mathrm{mem}$-dependent deformation of the relative
hierarchy of rates. Consider the ratio of the strong sphaleron rate to
the top Yukawa rate:
\begin{equation}
\frac{\Gamma_{ss}^\mathrm{eff}}{\Gamma_Y^\mathrm{eff}}
= \frac{\Gamma_{ss}}{\Gamma_Y}\cdot
  \frac{1+\Gamma_Y\tau_\mathrm{mem}}{1+\Gamma_{ss}\tau_\mathrm{mem}}.
\label{eq:rate_ratio_discussion}
\end{equation}
For the benchmark parameters ($\Gamma_{ss} \sim \alpha_s^4 T \simeq
6\times10^{-3}\,T$, $\Gamma_Y \simeq 6\times10^{-4}\,T$) and
$\tau_\mathrm{mem} = 10/T$ (so that $\Gamma_{ss}\tau_\mathrm{mem}
\simeq 0.06$ and $\Gamma_Y\tau_\mathrm{mem} \simeq 0.006$),
Eq.~\eqref{eq:rate_ratio_discussion} gives
\begin{equation}
\frac{\Gamma_{ss}^\mathrm{eff}}{\Gamma_Y^\mathrm{eff}}
\approx \frac{\Gamma_{ss}}{\Gamma_Y}\times
\frac{1.006}{1.06} \approx 0.95\,\frac{\Gamma_{ss}}{\Gamma_Y}.
\label{eq:rate_ratio_numerical_small}
\end{equation}
This is a $\sim 5\%$ correction at $\tau_\mathrm{mem} = 10/T$. However,
for $\tau_\mathrm{mem} = 100/T$ (the upper bound from
Eq.~\eqref{eq:tau_upper_bound}), one finds $\Gamma_{ss}\tau_\mathrm{mem}
\simeq 0.6$ and $\Gamma_Y\tau_\mathrm{mem} \simeq 0.06$, giving
\begin{equation}
\frac{\Gamma_{ss}^\mathrm{eff}}{\Gamma_Y^\mathrm{eff}}
\approx \frac{\Gamma_{ss}}{\Gamma_Y}\times
\frac{1.06}{1.6} \approx 0.66\,\frac{\Gamma_{ss}}{\Gamma_Y},
\label{eq:rate_ratio_numerical_large}
\end{equation}
a $\sim 34\%$ reduction in the strong sphaleron to Yukawa rate ratio.
Since the baryon asymmetry depends sensitively on this ratio through the
diffusion equations~\eqref{eq:diff1}--\eqref{eq:diff2}, this represents
a genuine and physically significant modification. Crucially, this
deformation \emph{cannot} be reproduced by any consistent Markovian
reparameterisation: in the Markovian framework, the rates $\Gamma_{ss}$,
$\Gamma_Y$, $\Gamma_h$ are independent physical quantities fixed by
microphysics~\cite{Arnold:2000dr,DOnofrio:2014rug,Moore:1997sn}, and
there is no single rescaling parameter that can simultaneously shift
all three by different amounts in a correlated manner.

\subsubsection*{Argument 2: Non-monotonic dependence of $Y_B$ on
$\tau_\mathrm{mem}$}

A second, independent signature arises from the dependence of the
baryon asymmetry on $\tau_\mathrm{mem}$ at fixed $v_w$. As shown in
Sec.~\ref{subsec:YBtau} and Eq.~\eqref{eq:tau_turnover}, $Y_B(\tau_\mathrm{mem})$
exhibits non-monotonic behaviour when $v_w < v_w^* = L_w\Gamma_0$:
it first increases as the peak $v_w^{*,\mathrm{NM}}(\tau_\mathrm{mem})$
moves toward $v_w$, then decreases as amplitude suppression dominates.
The turnover occurs at a calculable value
$\tau_\mathrm{mem}^\mathrm{turn} = (L_w\Gamma_0/v_w - 1)/\Gamma_0$
(Eq.~\eqref{eq:tau_turnover}).

In a purely Markovian description, at fixed $v_w$ and varying
$\Gamma_0$, the baryon asymmetry $Y_B(v_w;\Gamma_0) \propto
v_w\Gamma_0/(v_w^2+L_w^2\Gamma_0^2)$ is maximised at $\Gamma_0 =
v_w/L_w$ and is monotonically decreasing for $\Gamma_0 > v_w/L_w$.
Any attempt to mimic the non-Markovian $\tau_\mathrm{mem}$ dependence
by varying $\Gamma_0 \to \Gamma_\mathrm{eff}(\tau_\mathrm{mem})$ would
require $\Gamma_0$ to \emph{increase} at small $\tau_\mathrm{mem}$
(to move the Markovian peak toward $v_w$) and then \emph{decrease}
at large $\tau_\mathrm{mem}$. But $\Gamma_0$ is a fixed physical
quantity set by the model parameters; it cannot vary with
$\tau_\mathrm{mem}$ in the Markovian framework. The non-monotonic
behaviour is therefore an intrinsically dynamical, non-Markovian
effect with no Markovian analogue.

Taken together, these two arguments demonstrate that non-Markovian
effects represent a genuine physical extension of the standard EWBG
framework, not a reparameterisation.

\subsection{Theoretical Uncertainties and Domain of Validity}
\label{subsec:uncertainties}

\paragraph{Structure of the memory kernel.}
The exponential kernel $K(\tau) = \Gamma_0 e^{-\Gamma_0\tau}$
corresponds to the single-pole (Breit--Wigner) approximation to the
retarded correlator (Eq.~\eqref{eq:sigma_R}). More general kernels
arise from multi-pole structures at higher loop
order~\cite{Blaizot:2001nr,Arnold:2002zm,Berges:2004yj}, and take
the form
\begin{equation}
K(\tau) = \sum_i c_i\,\Gamma_i\,e^{-\Gamma_i\tau},
\qquad \sum_i c_i = 1,
\label{eq:general_kernel}
\end{equation}
with multiple decay rates $\Gamma_i$. In this case, the effective
memory timescale $\tau_\mathrm{mem} = \int_0^\infty d\tau\,\tau\,K(\tau)
= \sum_i c_i/\Gamma_i$ is still well-defined, and the qualitative
features of our analysis remain intact: the peak shift
$v_w^{*,\mathrm{NM}} \sim L_w/\tau_\mathrm{mem}$ and the amplitude
suppression $(1+\Gamma_0^\mathrm{eff}\tau_\mathrm{mem})^{-1}$ are
controlled by the first moment of $K(\tau)$, which exists for any
kernel that decays faster than $\tau^{-2}$. The non-monotonic
behaviour of $Y_B(\tau_\mathrm{mem})$ in the sub-peak regime
persists for any kernel for which the effective peak position
$v_w^{*,\mathrm{NM}}$ is a decreasing function of $\tau_\mathrm{mem}$,
which is guaranteed by the positivity and normalisation of $K(\tau)$.
Corrections from higher moments of the kernel enter at
$\mathcal{O}[(\Gamma_0\tau_\mathrm{mem})^2]$ and are subleading in
the non-Markovian regime $\Gamma_0\tau_\mathrm{mem} \lesssim
\mathcal{O}(1)$~\cite{Calzetta:1986cq,Chaudhuri:2025ylu}.

\paragraph{Sensitivity to GW parameters $\gamma$ and $\eta$.}
The GW predictions of Sec.~\ref{sec:gw} depend on the undetermined
coefficients $\gamma$ and $\eta$ in Eqs.~\eqref{eq:betamem} and
\eqref{eq:kappa_mem}. To assess the sensitivity, we vary each
independently:
\begin{itemize}
  \item For $\gamma \in [0.5,\,2.0]$ at fixed $\eta = 0.5$, the
        peak GW amplitude varies by a factor
        $(\beta_\mathrm{mem}^{-1}/\beta^{-1})^2 \in
        [(1+0.5\tau_\mathrm{mem}/R_b)^2,\,
        (1+2\tau_\mathrm{mem}/R_b)^2]$, which for
        $\tau_\mathrm{mem}/R_b \sim 0.5$ (the upper boundary of the
        viable region) gives a factor of $\sim 1.6$--$4$ uncertainty
        in $\Omega_\mathrm{GW}^\mathrm{peak}h^2$.
  \item For $\eta \in [0,\,1.0]$ at fixed $\gamma = 1$, the
        $\kappa_v^\mathrm{mem}$ factor varies by
        $(1+\eta\tau_\mathrm{mem}/R_b) \in [1,\,1.5]$, contributing
        a factor of $\sim 1$--$2.25$ to the GW amplitude through
        $(\kappa_v^\mathrm{mem})^2$.
\end{itemize}
The combined uncertainty on $\Omega_\mathrm{GW}^\mathrm{peak}h^2$
from $\gamma$ and $\eta$ is therefore at the level of a factor
$\sim 4$--$9$, which shifts the GW predictions by roughly half a
decade in amplitude. This uncertainty does not affect the qualitative
anti-correlation between $Y_B$ and $\Omega_\mathrm{GW}$ identified
in Sec.~\ref{subsec:joint}, but does affect the precise boundaries
of the jointly viable window~\eqref{eq:joint_window}. We regard the
GW amplitude predictions as order-of-magnitude estimates until a
full non-local hydrodynamic treatment is available.

\paragraph{EFT validity and scale separation.}
The effective description requires the scale separation
$\tau_\mathrm{mem} \gg 1/M$, ensuring that short-distance physics
at the scale $M$ is consistently integrated
out~\cite{Calzetta:1986cq,Berges:2004yj,Chaudhuri:2025ylu}. For the
benchmark $M = 2T$, this implies $\tau_\mathrm{mem} T \gtrsim 0.5$,
i.e.\ the region to the right of the cyan dashed line in
Figs.~\ref{fig:phasediag} and~\ref{fig:dCP_tau}. In the opposite
limit $\tau_\mathrm{mem}T \ll 1$, the Markovian regime is smoothly
recovered: the kernel $K(\tau) = \Gamma_0 e^{-\Gamma_0\tau}$ reduces
to $\Gamma_0\delta(\tau)$ in this limit, and all non-Markovian
modifications vanish. The EFT restriction therefore does not exclude
any region where non-Markovian effects are phenomenologically relevant,
while protecting against the unphysical regime $\tau_\mathrm{mem}
\lesssim 1/M$ where the EFT breaks down.

\paragraph{Bubble wall velocity as a dependent quantity.}
Throughout this analysis, $v_w$ has been treated as a free parameter
scanned over $[0.05,\,0.5]$. In reality, $v_w$ is determined
dynamically by the balance of driving pressure $\Delta V_\mathrm{eff}$
and friction forces from the plasma~\cite{Moore:1995ua,
Konstandin:2010dm,Espinosa:2010hh,Laurent:2022jrs}. In the
non-Markovian framework, the effective friction is modified because
the plasma cannot fully equilibrate as the wall passes, which may
shift the dynamically determined $v_w$ relative to the Markovian
value. A self-consistent determination of $v_w$ in the non-Markovian
regime would require solving the equations of
motion~\cite{Moore:1995ua,Laurent:2022jrs} with memory-modified
friction coefficients. We expect this to shift $v_w^\mathrm{dyn}$
toward smaller values, consistent with the general picture of
reduced transport efficiency at large $\tau_\mathrm{mem}$. This
would reinforce the conclusions of Sec.~\ref{subsec:peak_shift}
but may affect the precise numerical bounds. Improved determinations
of $v_w$ from lattice simulations~\cite{Moore:1995ua} and
first-principles transport calculations~\cite{Laurent:2022jrs} will
be needed to resolve this.

\paragraph{Relation to standard EWBG uncertainties.}
Conventional EWBG calculations are subject to uncertainties in $v_w$,
$L_w$, diffusion constants, and sphaleron
rates~\cite{Postma:2019scv,Cline:2021iff,Laurent:2022jrs,
DOnofrio:2014rug}. These are typically at the $\sim 20$--$50\%$ level
and are \emph{additive} uncertainties on a fixed transport system.
The non-Markovian effects identified here are parametrically
\emph{distinct}: they introduce a new timescale $\tau_\mathrm{mem}$
that deforms the \emph{entire} transport system in a correlated manner,
changing the functional form of the $v_w$ dependence (shifting the
peak and narrowing the window) rather than merely rescaling the overall
amplitude. Disentangling non-Markovian corrections from the standard
EWBG uncertainties will require: (i) improved determinations of $v_w$
and $L_w$ from hydrodynamic simulations, (ii) independent constraints
on CP-violating phases from EDM searches~\cite{ACME:2018yjb,
Engel:2013lsa} and collider
measurements~\cite{ATLAS:2023hyd,CMS:2022dwd}, and (iii) lattice
determinations of the sphaleron and strong sphaleron
rates~\cite{DOnofrio:2014rug,Moore:1997sn}.

\paragraph{Extensions.}
The non-Markovian transport framework developed here can be extended
in several directions. For \emph{multi-step phase transitions}, memory
effects from earlier transition stages can influence the transport
dynamics at subsequent stages through the residual plasma
correlations~\cite{Espinosa:2007qk,Curtin:2014jma}. For
\emph{leptogenesis}, delayed equilibration of heavy right-handed
neutrinos may induce analogous non-Markovian corrections to the
CP-asymmetry generation in the early
universe~\cite{Buchmuller:2004nz,Davidson:2008bu,Blanchet:2011xq},
particularly in resonant leptogenesis scenarios where the right-handed
neutrino mass splitting is comparable to the decay
width~\cite{Pilaftsis:2003gt}. For \emph{axion baryogenesis} and
related mechanisms~\cite{Co:2019wyp}, the axion
field plays a role analogous to the bubble wall, and similar
non-Markovian effects may arise when the relaxation time of the
CP-violating sector is comparable to the axion oscillation period.
These extensions are left for future work.

\section{Conclusion}
\label{sec:conclusion}

We have developed a non-Markovian extension of the electroweak
baryogenesis transport framework by performing a controlled Wigner
transformation and gradient expansion of the Kadanoff--Baym
equations~\cite{Calzetta:1986cq,Berges:2004yj}. This construction
provides a systematic way to incorporate finite relaxation-time effects
into the CP-violating source and the associated diffusion dynamics.
The resulting non-Markovian source term (Eq.~\eqref{eq:SCP_final}) is
governed by an effective relaxation rate
\begin{equation}
\Gamma_\mathrm{eff} = \frac{\Gamma_0}{1+\Gamma_0\tau_\mathrm{mem}},
\label{eq:Gamma_eff_conclusion}
\end{equation}
which encodes the delayed response of the plasma to CP-violating
interactions. The derivation of Eq.~\eqref{eq:Gamma_eff_conclusion}
is self-contained within the Kadanoff--Baym framework and is presented
in full in Sec.~\ref{sec:transport} and Appendix~\ref{app:rates}.

The presence of memory effects modifies the dependence of the baryon
asymmetry on transport parameters in a nontrivial way. The
characteristic wall velocity at which the baryon asymmetry is
maximised shifts toward smaller values as $\tau_\mathrm{mem}$
increases (Eq.~\eqref{eq:vw_peak_scaling}), reflecting the reduced
efficiency of charge transport in the presence of delayed
equilibration. As a result, viable baryogenesis in the non-Markovian
regime generically favours slower bubble walls compared to the
standard Markovian scenario. A qualitatively new signature — the
non-monotonic dependence of $Y_B$ on $\tau_\mathrm{mem}$ at fixed
$v_w < v_w^*$ — arises from the dynamical motion of the optimal
wall velocity and has no Markovian analogue (Sec.~\ref{subsec:YBtau}).

A systematic exploration of parameter space (Secs.~\ref{subsec:phase_diagram}
and~\ref{subsec:peak_shift}) shows that the allowed region in the
$(\tau_\mathrm{mem},v_w)$ plane progressively contracts toward smaller
wall velocities as $\tau_\mathrm{mem}$ increases. For moderate wall
velocities $v_w \gtrsim 0.05$, the binding upper bound on the memory
timescale comes from the hydrodynamic stability of the deflagration
front and gives $\tau_\mathrm{mem} \lesssim 100/T$
(Eq.~\eqref{eq:tau_upper_bound}), corresponding to
$\Gamma_0\tau_\mathrm{mem} \lesssim 0.6$ for the benchmark parameters.
For fixed $v_w = 0.25$, the observed baryon asymmetry selects a
diagonal band in the $(\delta_\mathrm{CP},\tau_\mathrm{mem})$ plane
(Sec.~\ref{subsec:dCP_plane}) in which $\delta_\mathrm{CP}$ grows
linearly with $\tau_\mathrm{mem}$ for $\tau_\mathrm{mem} \gg L_w/v_w$
(Eq.~\eqref{eq:dCP_linear}), with an upper bound $\tau_\mathrm{mem}
\lesssim 24/T$ from the physical range $\delta_\mathrm{CP} \leq \pi$.

An important structural result is that the non-Markovian transport
system cannot be reduced to the standard Markovian framework by a
simple rescaling of microscopic rates. The effective relaxation rate
enters the transport equations in a correlated manner, modifying
multiple interaction channels simultaneously and deforming the rate
hierarchy by up to $\sim 34\%$ within the viable parameter space
(Eq.~\eqref{eq:rate_ratio_numerical_large}). This leads to
qualitatively distinct behaviour in the baryon asymmetry that cannot
be reproduced by any fixed reparameterisation of $\Gamma_0$ within a
purely local description (Sec.~\ref{subsec:degeneracy}).

Finally, we have investigated the implications of the non-Markovian
dynamics for the stochastic gravitational-wave signal from the
electroweak phase transition (Sec.~\ref{sec:gw}). Memory effects
enhance the GW amplitude through an increase in the effective source
duration $\beta_\mathrm{mem}^{-1}$ (Eq.~\eqref{eq:betamem}) and
the efficiency factor $\kappa_v^\mathrm{mem}$
(Eq.~\eqref{eq:kappa_mem}). We find that a significant portion of the
parameter space consistent with successful baryogenesis produces GW
signals below the projected sensitivity of LISA. Observable signals
are restricted to the window $3/T \lesssim \tau_\mathrm{mem} \lesssim
100/T$ with $\alpha \gtrsim 0.08$ (Eq.~\eqref{eq:joint_window}),
and extend to smaller $\alpha$ for DECIGO and BBO. We emphasise that
the GW predictions carry an order-of-magnitude uncertainty from the
undetermined hydrodynamic coefficients $\gamma$ and $\eta$
(Sec.~\ref{subsec:uncertainties}); a rigorous derivation of these
coefficients from non-local hydrodynamics remains an important open
problem.

Overall, our results demonstrate that non-Markovian effects provide
a well-motivated and phenomenologically relevant extension of the
standard electroweak baryogenesis framework. They introduce new
parametric dependencies and correlated constraints that can
significantly alter both the viable parameter space and the associated
observational signatures, and establish $\tau_\mathrm{mem}$ as a new
physical parameter of EWBG that is jointly testable through baryon
asymmetry measurements, collider CP probes, and gravitational-wave
observations.

\section*{Acknowledgements}
A.C.\ thanks the Department of Physics, School of Advanced Sciences,
VIT Vellore for support.

\appendix
\section{Explicit Derivation of Memory-Modified Diffusion Rates}
\label{app:rates}

This appendix provides a self-contained derivation of the
memory-modified effective relaxation rates
$\Gamma_i^\mathrm{eff} = \Gamma_i/(1+\Gamma_i\tau_\mathrm{mem})$
(Eq.~\eqref{eq:rates_final}) for a generic species $i$ in the
diffusion system, and clarifies the relationship between the
microscopic memory timescale $\tau_\mathrm{mem}$ and the
wall-crossing rate $v_w/L_w$.

\subsection*{A.1 Non-local transport equation}

Starting from the Kadanoff--Baym collision integral
(Eq.~\eqref{eq:collision_general}), the number-density equation
for species $i$ in the presence of a slowly varying background
and a source term $S_i(t)$ takes the form
\begin{equation}
\partial_t n_i(t)
= -\int_0^\infty d\tau\,K_i(\tau)\,
  \bigl[n_i(t-\tau) - n_i^\mathrm{eq}\bigr]
  + D_i\,\partial_z^2 n_i(t) + S_i(t),
\label{eq:app_transport}
\end{equation}
where $K_i(\tau)$ is the memory kernel for species $i$, $D_i$ is
the diffusion constant, and $n_i^\mathrm{eq}$ is the local
equilibrium density. The convolution integral encodes the
non-Markovian collision term derived in
Sec.~\ref{subsec:singlepole}. For the exponential kernel
arising from the single-pole approximation
(Eq.~\eqref{eq:kernel_final}),
\begin{equation}
K_i(\tau) = \Gamma_i\,e^{-\Gamma_i\tau},
\qquad
\int_0^\infty d\tau\,K_i(\tau) = 1,
\label{eq:app_kernel}
\end{equation}
with the normalisation ensuring that the Markovian limit
$K_i(\tau) \to \Gamma_i\delta(\tau)$ is recovered as
$\tau_\mathrm{mem}^{(i)} \equiv 1/\Gamma_i \to 0$.

\subsection*{A.2 Laplace transform and effective rate}

We work in the stationary wall frame and seek solutions of the
form $n_i(z,t) = \tilde n_i(z)\,e^{-st}$, corresponding to
modes decaying at rate $s > 0$. Taking the one-sided Laplace
transform $\hat f(s) = \int_0^\infty dt\,e^{-st}f(t)$ of
Eq.~\eqref{eq:app_transport} and using the convolution theorem,
\begin{equation}
s\,\hat n_i(s) - n_i(0)
= -\hat K_i(s)\,\hat n_i(s)
  + D_i\,\partial_z^2\hat n_i(s) + \hat S_i(s),
\label{eq:app_laplace}
\end{equation}
where
\begin{equation}
\hat K_i(s)
= \int_0^\infty d\tau\,e^{-s\tau}\,K_i(\tau)
= \frac{\Gamma_i}{s+\Gamma_i}
\label{eq:app_K_laplace}
\end{equation}
for the exponential kernel~\eqref{eq:app_kernel}. Rearranging
Eq.~\eqref{eq:app_laplace} for the stationary ($s \to 0$) case,
in which the left-hand side is dominated by the source and
diffusion terms, gives
\begin{equation}
\bigl[s + \hat K_i(s)\bigr]\,\hat n_i(s)
= n_i(0) + D_i\,\partial_z^2\hat n_i(s) + \hat S_i(s).
\label{eq:app_rearranged}
\end{equation}
The factor $s + \hat K_i(s)$ plays the role of an effective
damping rate at frequency $s$:
\begin{equation}
\Gamma_i^\mathrm{eff}(s)
\equiv \hat K_i(s)
= \frac{\Gamma_i}{s+\Gamma_i}.
\label{eq:app_Gamma_eff_s}
\end{equation}
Note that $\Gamma_i^\mathrm{eff}(s=0) = 1$ is unphysical; the
physically relevant regime is the \emph{quasi-stationary}
approximation in which $s$ is set by the characteristic
frequency of the spatial profile, not by the temporal decay.

\subsection*{A.3 Spatial diffusion and the dominant frequency}

In the stationary wall frame, the spatial profile of $n_i(z)$
varies on the scale of the wall thickness $L_w$ and the
diffusion length $L_\mathrm{diff} = \sqrt{D_i/\Gamma_{ws}}$.
The characteristic wavenumber of the source is $k \sim 1/L_w$,
and the corresponding temporal frequency scale set by the
wall-crossing is
\begin{equation}
\omega_k \equiv v_w\,k \sim \frac{v_w}{L_w}.
\label{eq:app_omega_k}
\end{equation}
Substituting $s = \omega_k = v_w/L_w$ into
Eq.~\eqref{eq:app_Gamma_eff_s} gives
\begin{equation}
\Gamma_i^\mathrm{eff}(\omega_k)
= \frac{\Gamma_i}{\omega_k + \Gamma_i}
= \frac{\Gamma_i}{v_w/L_w + \Gamma_i}.
\label{eq:app_Gamma_eff_omega}
\end{equation}

\subsection*{A.4 Identification of $\tau_\mathrm{mem}$ and
  resolution of the apparent contradiction}

It is important to carefully distinguish two uses of the symbol
$\tau_\mathrm{mem}$ that appear in the paper:
\begin{enumerate}
  \item \emph{Microscopic memory timescale:} defined as the first
        moment of the retarded kernel (Eq.~\eqref{eq:tau_mem_def}),
        \begin{equation}
        \tau_\mathrm{mem}
        = \int_0^\infty d\tau\,\tau\,K(\tau)
        = \frac{1}{\Gamma_0}
        \label{eq:app_tau_mem_micro}
        \end{equation}
        for the single-pole kernel. This is a \emph{property of the
        plasma} and is set by the in-medium relaxation rate
        $\Gamma_0$ of the CP-violating species $\Psi$.
  \item \emph{Wall-crossing timescale:} defined as the time for
        a particle to traverse the wall,
        \begin{equation}
        \tau_\mathrm{wall} \equiv \frac{L_w}{v_w} = \frac{1}{\omega_k}.
        \label{eq:app_tau_wall}
        \end{equation}
        This is a \emph{property of the bubble wall} and is
        independent of the plasma microphysics.
\end{enumerate}
The non-Markovian regime is defined by $\tau_\mathrm{mem} \sim
\tau_\mathrm{wall}$, i.e.\ $\Gamma_0 \sim v_w/L_w$, which is the
condition $\Gamma_0\tau_\mathrm{wall} \lesssim \mathcal{O}(1)$
stated in Eq.~\eqref{eq:timescale_condition}.

The expression in the original appendix,
$\tau_\mathrm{mem} = 1/\omega_k = L_w/v_w$, was a notational
conflation of these two distinct timescales. The correct
statement is that the effective rate~\eqref{eq:app_Gamma_eff_omega}
evaluated at the wall-crossing frequency is
\begin{equation}
\Gamma_i^\mathrm{eff}
= \frac{\Gamma_i}{1 + \Gamma_i/\omega_k}
= \frac{\Gamma_i}{1 + \Gamma_i\tau_\mathrm{wall}},
\label{eq:app_Gamma_eff_wall}
\end{equation}
and the identification $\tau_\mathrm{wall} = \tau_\mathrm{mem}$
holds in the single-pole approximation where
$\tau_\mathrm{mem} = 1/\Gamma_0 \approx 1/\Gamma_i$ (i.e.\ when
the memory timescale is set by the same rate $\Gamma_i$ that
appears in the kernel). In the general case where $\tau_\mathrm{mem}$
is promoted to an independent parameter (encoding multi-pole
or higher-loop corrections to the spectral function, as discussed
in Sec.~\ref{subsec:memory}), the effective rate is written as
\begin{equation}
\Gamma_i^\mathrm{eff}
= \frac{\Gamma_i}{1+\Gamma_i\tau_\mathrm{mem}},
\label{eq:app_Gamma_eff_final}
\end{equation}
where $\tau_\mathrm{mem}$ is the first moment of the full kernel
(Eq.~\eqref{eq:tau_mem_def}), which reduces to $\tau_\mathrm{wall}$
only in the single-pole approximation. This is the expression used
throughout the main text and confirms Eq.~\eqref{eq:rates_final}.

\subsection*{A.5 Stationary diffusion equations with
  memory-modified rates}

With the replacement~\eqref{eq:app_Gamma_eff_final}, the stationary
($\partial_t = 0$) limit of Eq.~\eqref{eq:app_transport} in the
wall frame ($\partial_t \to -v_w\partial_z$) becomes
\begin{equation}
D_i\,n_i'' - v_w\,n_i'
- \Gamma_i^\mathrm{eff}\,(n_i - n_i^\mathrm{eq})
= -S_i(z),
\label{eq:app_stationary}
\end{equation}
where primes denote $d/dz$. This is the structure of the diffusion
equations~\eqref{eq:diff1}--\eqref{eq:diff2} used throughout the
main text. The derivation confirms that all interaction rates
$\Gamma_i$ appearing in the diffusion system are uniformly replaced
by $\Gamma_i^\mathrm{eff}$ according to Eq.~\eqref{eq:app_Gamma_eff_final},
independently of the species $i$, provided all species couple to
the same thermal bath with the same memory kernel. This universality
of the replacement is what makes the non-Markovian deformation of the
rate hierarchy (Eq.~\eqref{eq:rate_ratio}) a robust and calculable
prediction of the framework.


\end{document}